\documentclass[fleqn,usenatbib]{mnras}

\usepackage{newtxtext,newtxmath}

\usepackage[T1]{fontenc}
\usepackage{subcaption}

\DeclareRobustCommand{\VAN}[3]{#2}
\let\VANthebibliography\thebibliography
\def\thebibliography{\DeclareRobustCommand{\VAN}[3]{##3}\VANthebibliography}

\usepackage{graphicx}
\usepackage{rotating}
\usepackage{multirow}
\usepackage{booktabs}
\usepackage{xcolor}
\usepackage{xspace}

\newcommand{\nustar}{{\it NuSTAR}\xspace}
\newcommand{\nicer}{{\it NICER}\xspace}
\newcommand{\swift}{{\it Swift}\xspace}

\newcommand{\astrosat}{{\it AstroSat}\xspace}
\newcommand{\maxi}{{\it MAXI}\xspace}
\newcommand{\ergs}{erg s$^{-1}$\xspace }

\newcommand\T{\rule{0pt}{2.6ex}}       
\newcommand\B{\rule[-1.2ex]{0pt}{0pt}}
\newcommand{\ixpe}{{\it IXPE}\xspace}
\newcommand{\source}{4U 1820-30\xspace}

\title[The accretion/ejection link in the neutron star X-ray binary 4U 1820-30]{The accretion/ejection link in the neutron star X-ray binary 4U 1820-30 I: \\ A boundary layer-jet coupling?}

\author[A. Marino]{
A. Marino$^{1,2,3}$\thanks{E-mail: marino@ice.csic.es}, T. D. Russell$^{3}$, M. Del Santo$^{3}$, A. Beri$^{4,5}$, A. Sanna$^{6}$, F. Coti Zelati$^{1,2}$, N. Degenaar$^{7}$
\newauthor
D. Altamirano$^{5}$, E. Ambrosi$^{3}$, A. Anitra$^{8}$, F. Carotenuto$^{9}$, A. D'A\`i$^{3}$, T. Di Salvo$^{8}$, A. Manca$^{6}$, S. E. Motta$^{10}$
\newauthor
C. Pinto$^{3}$, F. Pintore$^{3}$, N. Rea$^{1,2}$, J. van den Eijnden$^{11}$
\\
$^{1}$Institute of Space Sciences (ICE, CSIC), Campus UAB, Carrer de Can Magrans s/n, E-08193 Barcelona, Spain \\
$^{2}$Institut d'Estudis Espacials de Catalunya (IEEC), Carrer Gran Capit\`a 2--4, E-08034 Barcelona, Spain \\
$^{3}$INAF/IASF Palermo, via Ugo La Malfa 153, I-90146 - Palermo, Italy \\
$^{4}$Indian Institute of Science Education and Research (IISER) Mohali, Punjab 140306, India \\
$^{5}$School of Physics and Astronomy, University of Southampton, Southampton, SO17 1BJ, UK \\
$^{6}$Universit\`a degli Studi di Cagliari, Dipartimento di Fisica, SP Monserrato-Sestu km 0.7, I-09042 Monserrato, Italy \\
$^{7}$Anton Pannekoek Institute for Astronomy, University of Amsterdam, Science Park 904, 1098 XH, Amsterdam, the Netherlands\\
$^{8}$Universit\`a degli Studi di
  Palermo, Dipartimento di Fisica e Chimica, via Archirafi 36 - 90123 Palermo, Italy \\
$^{9}$Astrophysics, Department of Physics, University of Oxford, Keble Road, Oxford OX1 3RH, UK\\
$^{10}$Istituto Nazionale di Astrofisica, Osservatorio Astronomico di Brera, via E. Bianchi 46, I-23807 Merate (LC), Italy\\
$^{11}$Department of Physics, University of Warwick, Coventry CV4 7AL, UK\\
}

\date{Accepted XXX. Received YYY; in original form ZZZ}

\pubyear{2023}

\begin{document}
\label{firstpage}
\pagerange{\pageref{firstpage}--\pageref{lastpage}}
\maketitle

\begin{abstract}

The accretion flow / jet correlation in neutron star (NS) low-mass X-ray binaries (LMXBs) is far less understood when compared to black hole (BH) LMXBs. In this paper we will present the results of a dense multi-wavelength observational campaign on the NS LMXB 4U 1820-30, including X-ray (\nicer, \nustar and \astrosat) and quasi-simultaneous radio (ATCA) observations in 2022. \source shows a peculiar 170 day super-orbital accretion modulation, during which the system evolves between "modes" of high and low X-ray flux. During our monitoring, the source did not show any transition to a full hard state. X-ray spectra were well described using a disc blackbody, a Comptonisation spectrum along with a Fe K emission line at $\sim$6.6 keV. Our results show that the observed X-ray flux modulation is almost entirely produced by changes in the size of the region providing seed photons for the Comptonisation spectrum. This region is large ($\sim$15 km) in the high mode and likely coincides with the whole boundary layer, while it shrinks significantly ($\lesssim$10 km) in low mode. The electron temperature of the corona and the observed RMS variability in the hard X-rays also exhibit a slight increase in low mode. As the source moves from high to low mode, the radio emission due to the jet becomes $\sim$5 fainter. These radio changes appear not to be strongly connected to the hard-to-soft transitions as in BH systems, while they seem to be connected mostly to variations observed in the boundary layer.

\end{abstract}

\begin{keywords}
accretion, accretion discs -- stars:neutron -- X-rays: binaries -- X-rays, individuals: 4U 1820-30 -- ISM: jets and outflows
\end{keywords}



\section{Introduction}
Neutron Star (NS) low-mass X-ray Binaries (LMXBs) are binary systems composed of a NS accreting matter from a companion low-mass, i.e., typically lower than 1 M$_\odot$, star. The material accreted onto the compact object is responsible for most of the electromagnetic emission of these sources from the optical to the X-rays. These systems have been historically grouped in two main classes, Z- and atoll sources, based on the particular tracks they draw in X-rays colour-colour diagrams \citep[][]{Hasinger1989}. \\ Z-sources are the brightest among the two classes, being persistently active and always accreting around the Eddington limit, i.e., at a typically observed X-ray luminosity range of L$_X\sim10^{37}-10^{38}$ erg/s \citep[see, e.g.][]{VanDerKlis2006}. Atolls are typically fainter, i.e., at L$_X\sim10^{36}-10^{37}$ erg/s \citep{MunozDarias2014} and thereby are considered to be accreting at lower rates. They can be persistent or transients, i.e. usually found in a dormant, faint X-ray regime called quiescence with only episodic outbursts where the X-ray luminosity becomes comparable to the level of the persistent NS LMXBs. However, the mere existence of several systems showing both Z or atoll behaviours make the distinction between these regimes somehow blurred. This is the case for XTE J1701-462 which transitioned from an atoll to a Z-source regime in both its 2006/2007 \citep[see, e.g.,][]{Homan2010} and 2022 \citep{Homan2022a,Homan2022b} outbursts and the other two transient Z-sources IGR J17480-2446 \citep[e.g.][]{Chakraborty2011} and MAXI J0556-332 \citep{Homan2011}. \\ The accretion flow consists typically of an optically thick accretion disc, an optically thin cloud of hot electrons usually dubbed corona, and a boundary layer (BL) connecting the inner edge of the disc with the NS surface. 
The observed X-ray spectral continuum from these sources can be broken down in several components, corresponding to these regions of the accretion flow: a multi-colour disc black body in the soft X-rays and a Comptonisation spectrum, which can extend up to hundreds of keV \citep{Pintore2016, Ludlam2016, Gambino2019}. The contribution from a boundary layer between the disc and the NS surface, or the surface itself, can also be visible directly as an additional black body component or as source of the seed photons for the Comptonisation spectrum at a blackbody temperature $kT_{\rm bb}$ of about 1-2 keV \citep[see, e.g.,][]{Mondal2020,Marino2022}. \\ The X-ray spectral-timing properties of these objects evolve between two main states: one dominated by the corona emission, typically referred to as the hard state, and the other dominated by the accretion disc and/or the NS/boundary layer, i.e., the soft state \citep[for a review,][]{Done2007}. \\ 
In the radio--mid-IR region, the contribution of a compact jet, i.e., a collimated ionised outflow of relativistic particles ejected by the system, is instead dominant. The observed properties of the compact jet show a clear correlation with the accretion flow. This connection is well-established for black hole (BH) transients \citep[see, e.g.,][]{Corbel2000,Coriat2011}, where as their X-ray spectral-timing properties display dazzling variations between hard and soft states, similar to the states observed in NS LMXBs, the jet evolves from being radio loud to a quenched state, respectively \citep{Fender2004}. These radio/X-ray changes occur usually over time-scales of the order of one or few weeks in both BH and NS LMXBs \citep[see, e.g.,][]{Russell2014,Marino2019b,Marcel2019,DeMarco2021,Rhodes2022}. However, this standard pattern is less clear for NS systems. Despite being observed several times \citep[e.g.][]{Gusinskaia2017}, it is yet not established whether jet quenching in the soft state is also the norm for all NS LMXBs, with some cases where the presence of compact jets was reported even after the transition to the soft state \citep[][]{Migliari2004}.  Interestingly, jet quenching in some accreting NSs has been sometimes observed without a state transition but corresponding to an evolution in the flux \citep{Russell2021, Panurach2023}, suggesting a more complex accretion/ejection coupling with respect to BH X-ray binaries (XRBs). When observed, jets in NS LMXBs are generally $\sim$20 times radio-fainter than BH LMXBs \citep[e.g., ][]{Gallo2018, VandenEijnden2021}, although several systems, often harbouring X-ray millisecond pulsars \citep[e.g.,][]{Russell2018,Gusinskaia2020,Cotizelati2021}, have been observed at radio luminosity comparable to BH systems. Different jet geometries and/or weaker couplings with the accretion flow in accreting NSs with respect to BHs have been also proposed \citep[][]{Marino2020}. The emerging observational picture is yet not well understood and, in particular, the role (if any) played by the NS magnetic field, its emitting surface, its spin, or the presence of a boundary layer is unclear. Multi-wavelength surveys of these systems that probe the behaviour of the jets and the accretion flow over a variety of mass accretion rates/X-ray states are crucial to self-consistently explore jet launching in NSs. \\

\subsection{4U 1820-303} \label{ss:source}
A particularly intriguing target to perform this kind of studies is the NS LMXB 4U 1820-30. This source is composed by a NS accreting matter from a white dwarf (WD) companion in a tight orbit. Indeed, with its orbital period of only 11.4 minutes \citep[][]{Stella1987}, i.e., the shortest orbital period known for any XRB, \source is identified as an Ultra-Compact XRB \citep[UCXB, for a recent review][]{ArmasPadilla2023}. Located at a distance of about 7.6 kpc \citep{Kuulkers2003}, the system resides within the globular cluster NGC 6624. The source is classified as an atoll and it is persistently accreting at a relatively high luminosity, displaying mostly soft X-ray spectral states (the so-called "banana" state for atolls), with occasional transitions to the hard state ("island" state) \citep{Tarana2007}. Its behaviour is indeed similar to what observed in the so-called "bright atolls" group \citep[e.g.,][]{Dai2010,Egron2013,Iaria2020}. \\ Perhaps the most peculiar aspect of the source is the $\sim$ 170 d, super-orbital accretion cycle exhibited by the source. The luminosity modulation is intrinsic, i.e., not due to absorption from occulting plasma \citep[][]{Zdziarski2007a} and it likely originates from the presence of a third object orbiting the system from afar \citep[][]{Chou2001}. According to this theory, the tidal forces exerted by this third object on \source trigger fluctuations in the orbital eccentricity of the binary, which in turn translate to periodic increases in the mass-accretion rate. As a consequence, \source\ oscillates between a low mode, at $L_{\rm low}\sim 8\times 10^{36}$ \ergs and a high mode, at $L_{\rm high}\sim 6\times 10^{37}$ \ergs (in the range 0.5-10 keV and assuming a distance D= 7.6 ${\rm kpc})$ over a time-scale of about 170 d. These modes can be characterized by different spectral properties, most likely connected to changes in the accretion flow. In the low mode, the system can be found in soft ("banana") and sometimes in the hard ("island") X-ray spectral states \citep{intzand2012}, during which it exhibits frequent type-I X-ray bursts. On the other hand, in the high mode, the system is only soft and bursts are typically absent \citep[see, e.g., ][]{Titarchuk2013}. \\ The presence of the Fe K line in the X-ray spectrum is debated, with detections reported only by a number of X-ray spectral studies \citep[][]{Cackett2008,Titarchuk2013} and not by others \citep[e.g., ][]{Costantini2012, Mondal2016}. 
Moreover, while \source\ was initially identified as a non-quenching jet NS \citep[][]{Migliari2004}, it has been recently suggested that compact jet quenching does occur \citep[][]{Russell2021}. Indeed, the change in spectral properties of the jet seem to occur only when, in the transition from low to high mode, the X-ray flux exceeds a certain threshold \citep[see, Fig. 2,][]{Russell2021}. Such a result would suggest a critical connection between jet and mass-accretion rate rather than spectral hardness, in a way that was never observed for BH systems and that is instead sometimes observed in accreting NSs, both in LMXBs \citep{Panurach2023} and in XRBs hosting high-mass companion stars \citep[HMXBs, see][]{VanDenEijnden2018_jet}. However, those results were obtained from observations taken sporadically over the past $\sim$17 years. To properly understand the accretion-ejection coupling in this system - and potentially in similar objects - it is fundamental to measure the accretion flow/jet properties over a single accretion cycle. \\
In this series of papers, we present the results of a comprehensive radio--X-rays observational campaign performed throughout 2022, aimed at following \source\ in the X-rays and radio during a whole super-orbital cycle. In this manuscript, we present the results of a broadband X-ray spectral analysis and timing analysis using \nicer\ , \nustar and \astrosat\ data from the 2022 April-September cycle. We also used data from the Australia Telescope Compact Array (ATCA) taken close to our broadband X-ray spectra to show the jet evolution during this cycle. A following paper (Paper II of this series, Russell et al., in preparation) will present the results of the complete radio monitoring and how they compare with the X-rays data analysis reported here.





\begin{figure*}
	\includegraphics[width=0.9\textwidth]{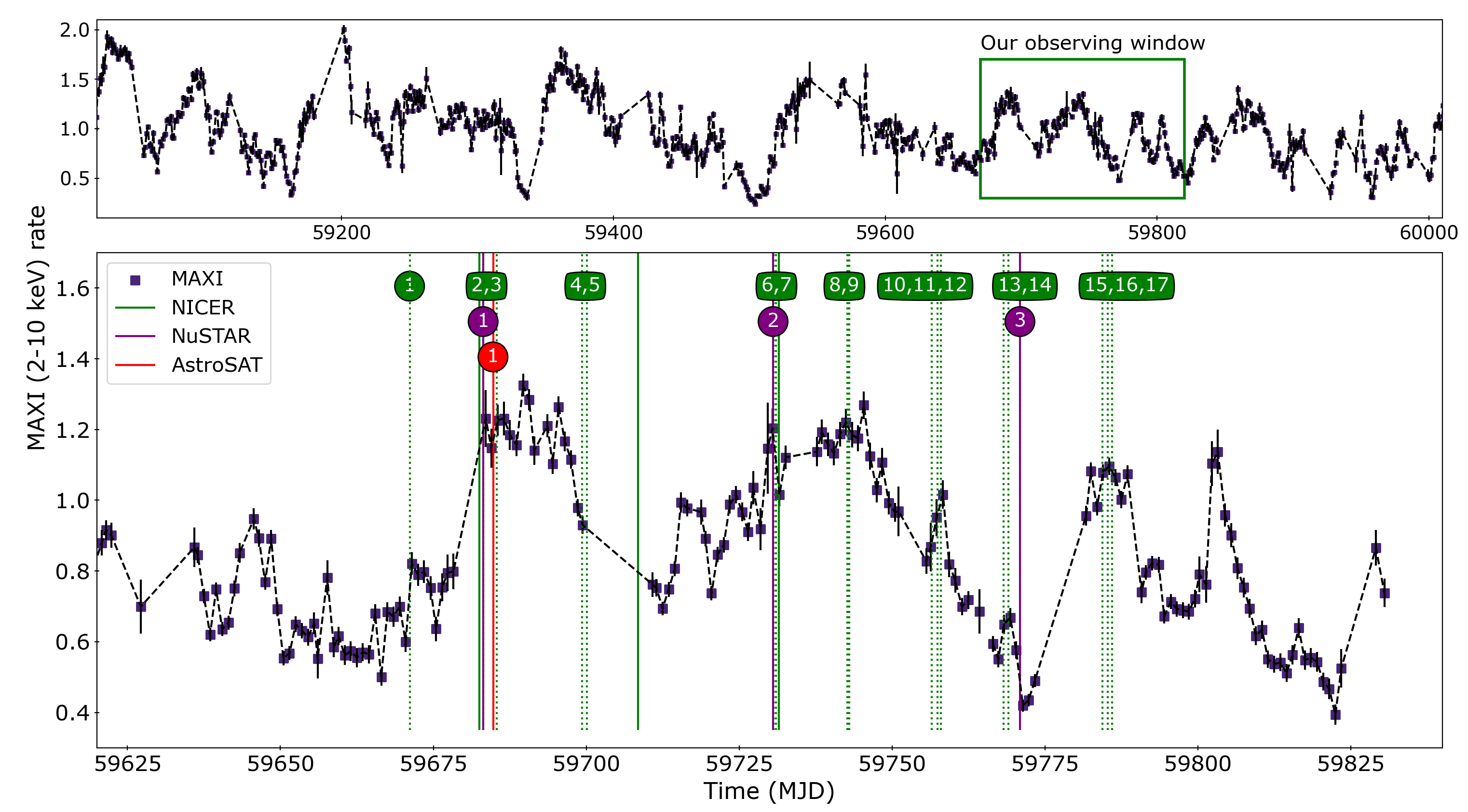}
    \caption{(\emph{Top}) \source daily light curve as observed by {\it MAXI} from June 2020 to June 2023, displaying the super-orbital modulation typical of the system. The time interval enclosing the times of the observations used in this paper is drawn as a green rectangle. (\emph{Bottom}) A zoom of the top light curve displaying \maxi light curve from April to September 2022. The observation dates for \nicer, \nustar and \astrosat are highlighted as vertical lines in green dashed, purple solid and red solid, respectively. Colored boxes with numbers are displayed to identify the dates of the single \nicer , \nustar and \astrosat observations using the labels introduced in Table \ref{tab:obs} and the same colors chosen for the vertical lines.}
    \label{fig:lcurve}
\end{figure*}

\section{Observations \& Data reduction}
A dense X-ray campaign has been performed between April 2 and July 26 2022, with 17, 3 and 1 observations for \nicer\ , \nustar\ and \astrosat\ , respectively. A summary of these observations is reported in Table \ref{tab:obs}. All data were reprocessed using the \textsc{HEAsoft} package v. 6.30. The latest available \textsc{CALDB} version was used for all the instruments. More details on the data reduction for each observatory are given in the following subsections.

\subsection{\nicer}
A two-weeks cadence monitoring campaign was performed by \nicer (see Table \ref{tab:obs}). Data were reduced using \texttt{nicerl2} task (\texttt{NICERDAS 2019-05-21 v006}). We set recommended calibration processes, standard screening and we added the \texttt{niprefilter2$\_$coltypes=base,3c50} parameter so that the \texttt{3C50} model can be used to derive background spectra later. We extracted the cleaned event files, checking that all detectors were active during observations and excluding data from the "noisy" detectors (labelled 14 and 34), in order to reduce the detector noise. We accumulated light curves from all the observations using the {\tt xselect} tool, with the aim to check for type-I X-ray bursts, finding only one in observation 5604010304. In the following data analysis, the burst was excluded. Then we selected the GTI using \texttt{NIMAKETIME} and applied them to the data via \texttt{NIEXTRACT-EVENTS}, selecting events with PI channel between 25 and 1200 (0.25–12.0~keV). We used the {\tt nibackgen3C50} tool to extract both the spectra of the source and the background from the cleaned event files, selecting the 2020 gain calibration. \\ 
During the writing of this paper, the updated \textsc{HEAsoft} version 6.31 was released, along with substantial changes to the \nicer mission data analysis pipeline. To check for consistency, we reanalysed three observations of the 2022 \nicer campaign using a different background model, i.e. the SCORPEON model. The spectra obtained with the different \textsc{HEAsoft} versions are basically identical and the best-fit parameters obtained are not significantly affected by the change of version and/or background model. We therefore did not reanalyse the whole data set and kept working on the spectra obtained with the previous \textsc{HEAsoft} version. 

\subsection{\nustar}
\nustar observed the system three times during this campaign, for a total exposure of about 58.6~ks. We reduced the data using standard tools within the \texttt{Nustardas} package. A 100" radius circular region centered at the coordinates of the source was selected as source region. In order to take into account any background non-uniformity on the detector, we extracted the background spectra using four circles of $\sim$50" radii placed on different areas which were well far away from the source region. Finally, we used \texttt{Nuproducts} to extract spectra and light curves. We systematically checked for the presence of type-I X-ray bursts within the \nustar observations, but we did not find any. 


\begin{table}
    \centering
    \begin{tabular}{ l l l l l }
         \hline
         \hline
         Id.$^a$ & ObsID & \multicolumn{2}{c}{Start Time} & Exposure  \T\\
         & & (UTC) & (MJD) & ks  \B\\
         \hline
         & \multicolumn{4}{c}{\nustar} \T \B \\
         \hline
         Nu01 & 30802009002 & 2022-04-14 & 59683 & 23.9  \\
         Nu02 & 30802009004 & 2022-05-31 & 59730 & 14.6  \\
         Nu03 & 30802009006 & 2022-07-10 & 59770 & 20.1  \\
         \hline
         & \multicolumn{4}{c}{\it AstroSat} \T \B \\
         \hline
         As01 & 9000005070 & 2022-04-15 & 59684 & 22.8 \\ 
         \hline
         & \multicolumn{4}{c}{ \nicer } \T \B \\
         \hline
         Ni01 & 5604010101 & 2022-04-02 & 59671 & 8.1 \\
         Ni02 & 5604010102 & 2022-04-14 & 59683 & 4.2 \\
         Ni03 & 5604010103 & 2022-04-16 & 59685 & 1.1 \\
         Ni04 & 5604010301 & 2022-04-30 & 59699 & 2.6 \\
         Ni05 & 5604010302 & 2022-05-01 & 59700 & 6.5 \\
         Ni06 & 5604010304 & 2022-05-31 & 59730 & 9.0 \\
         Ni07 & 5604010305 & 2022-06-01 & 59731 & 2.0 \\
         Ni08 & 5604010401 & 2022-06-12 & 59743 & 9.2 \\
         Ni09 & 5604010402 & 2022-06-13 & 59743 & 1.3 \\
        Ni10 & 5604010501 & 2022-06-26 & 59756 & 1.3 \\
        Ni11 & 5604010502 & 2022-06-27 & 59757 & 1.3 \\
        Ni12 & 5604010503 & 2022-06-27 & 59758 & 3.4 \\
        Ni13 & 5604010601 & 2022-07-08 & 59768 & 6.3 \\
        Ni14 & 5604010602 & 2022-07-09 & 59769 & 4.1 \\
        Ni15 & 5604010701 & 2022-07-24 & 59784 & 2.9 \\
        Ni16 & 5604010702 & 2022-07-25 & 59785 & 4.0 \\
        Ni17 & 5604010703 & 2022-07-26 & 59786 & 2.1 \\
        \hline
        \hline
    \end{tabular}
    \caption{List of the \nustar , \astrosat and \nicer observations of \source\ used in this work. $^a$Each observation is marked with an identification code.}
    \label{tab:obs}
\end{table}

\subsection{AstroSat}
As part of the multi-wavelength campaign, \astrosat observed \source on April 15, 2022.~We have analysed data from the Soft X-ray Telescope 
\citep[\textsc{SXT}, ][]{Singh2016} and the Large Area X-rays Proportional Counter \citep[\textsc{LAXPC}][]{Yadav2016, Antia2017, Antia2021} on-board \astrosat \citep{Agrawal2006, Singh2014}. 
\textsc{LAXPC}, one of the primary instruments on-board and consists of three co-aligned identical proportional counter detectors, viz. \textsc{LAXPC10}, \textsc{LAXPC20} and \textsc{LAXPC30}.~Each of these works in the energy range of 3$-$80~\rm{keV} \citep[for details see,][]{Yadav2016, Antia2017, Antia2021}.~Due to the gain instability caused by the gas leakage, \textsc{LAXPC10} data were not used while \textsc{LAXPC30} was switched off during these observations\footnote{LAXPC30 is switched off since 8 March 2018, refer to \url{http://astrosat-ssc.iucaa.in/}}.~Therefore, we have used data from \textsc{LAXPC20} for our work. These data were collected in the Event mode (EA) which contains the information about the time, channel number and anodeID of each event.~\textsc{LaxpcSoft} v3.3\footnote{\url{http://www.tifr.res.in/~astrosat\_laxpc/LaxpcSoft.html}} software package was used to extract spectra.
Background files are generated using the blank sky observations \citep[see][for details]{Antia2017}.


The \textsc{SXT} aboard \astrosat is a focusing instrument sensitive mainly in the 0.3–7~\rm{keV} energy band \citep{Singh2014,Singh2017} and its camera
assembly uses an e2v CCD, identical to that on
\emph{XMM-Newton}/\textsc{MOS} and \emph{Swift}-\textsc{XRT}. The \textsc{SXT} observations of \source were carried out in the photon counting mode. We have used Level-2 data provided by the \textsc{SXT} payload operation center (POC) in Mumbai, India, reduced using the most recent pipeline and calibration database (version 1.4b).~Events from each orbit were merged using \textsc{SXT} Event Merger Tool (Julia Code\footnote{\url{http://www.tifr.res.in/~astrosat\_sxt/dataanalysis.html}}).~These merged events were used to extract image and spectra using the ftool task \textsc{xselect}. In \textsc{SXT}, pile up effects are notable at count rates higher than 40~$cs^{-1}$\footnote{\url{https://web.iucaa.in/~astrosat/AstroSat_handbook.pdf}}. In our observations, the source showed 80~$cs^{-1}$, suggesting a significant pile-up.~To minimise the effect of the pile-up, source spectrum is extracted from an annulus between 5~$\arcmin$ and 15~$\arcmin$ from the centre of the image, following the method described in \citep{Chakraborty2020}.
We have used the response~(sxt\_pc\_mat\_g0to12.rmf) and background~(SkyBkg\_comb\_EL3p5\_Cl\_Rd16p0\_v01.pha) files provided by the \textsc{SXT} team.~The appropriate ARF file suitable for the specific source region is generated using the command line auxiliary tool \textsc{sxtARFModule}.


\begin{figure}
	\includegraphics[width=1.0\columnwidth]{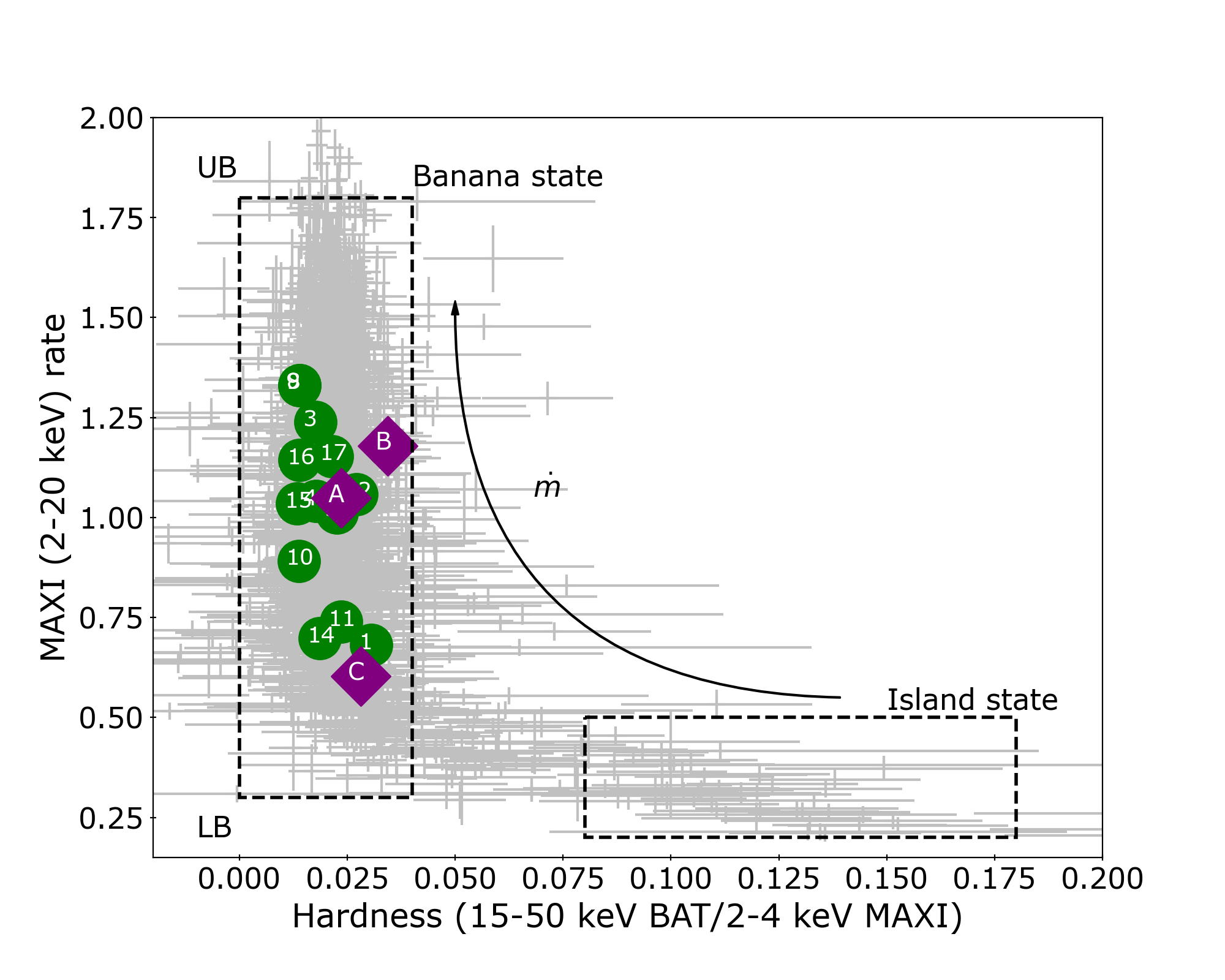}
    \caption{Hardness Intensity Diagram of \source , produced by using all available \maxi observations and all corresponding \swift /BAT count-rates. In particular we used \maxi count-rates over the 2--20 keV energy range for the intensity, while we estimate the hardness as the ratio between the BAT count-rate (15--50 keV) and the \maxi count-rate in the soft band (2--4 keV) taken the same day \citep[as in][]{Russell2021}. Superimposed to the grey data points, we highlight with green circles the position in the HID during the dates of the single \nicer observations, labeled with progressive numbers. For the \nicer observations used in broadband spectra A, B and C, we adopted purple diamonds with those letters as labels. The labels "UB" and "LB" indicate the locations of the "Upper" and "Lower" Banana branch.}
    \label{fig:HID}
\end{figure}
\section{Spectral analysis}\label{sec:spectral}
We present the three-year light curve of \source\ as observed by the Monitor of All Sky X-ray Image \citep[\maxi][]{Matsuoka2009} in Figure \ref{fig:lcurve} (top panel), displaying its super-orbital modulation. On the bottom panel, we zoom in on the period during which our campaign was carried out, with the dates of the \nicer\ , \nustar\ and \astrosat visits highlighted. It is apparent how our source showed a slightly more irregular behaviour during our $\sim$150 days observing window, with several highs and lows in rapid succession. In order to set the evolution of the system during this cycle in the framework of its observational history, we created a Hardness Intensity Diagram (HID) using \maxi and \swift\ /BAT data \citep{Krimm2013}. In particular we collected all the available \maxi\ count-rates (in the 2--4 keV and 2--20 keV energy ranges) since October 2009 and the corresponding, i.e. taken the same day, BAT count-rates (15--50 keV) from the respective websites\footnote{\maxi : \url{ http://maxi.riken.jp/top/index.html}, BAT: \url{https://swift.gsfc.nasa.gov/results/transients/}}. We then plotted \maxi total count-rates versus the ratio between BAT and \maxi count-rates in the 2--4 keV band to build the HID (see Fig. \ref{fig:HID}). In the Figure, we highlighted the position of the source in the HID in the days of the \nicer and/or \nustar observations, showing that the source lingered in the soft ("banana") state during the whole surveyed cycle. \\
In order to perform the spectral analysis, we grouped all \nicer\ and \nustar\ spectra through optimal binning \citep{Kaastra2016}, i.e. in order to have a grouping consistent with the spectral resolution of the instrument (and avoid oversampling) in a given energy range, but at the same time keeping at least 20 counts per bin to ensure the use of the $\chi^2$ statistics. The \textsc{SXT} and \textsc{LAXPC} spectra were grouped to have at least 25 counts/bin.~The \textsc{LAXPC} spectra showed a large calibration uncertainty, with background dominating above 20~keV.~Similar issues have also been observed in other observations \citep[see e.g.,][]{Beri2020, Sharma2023}. Therefore, we have used data only up to 20 keV for our spectral analysis. \\
We used \textsc{Xspec} v.12.12.1 for all the performed spectral fits. In all cases, we used the \textsc{tbabs} model to take into account interstellar absorption, including the photoelectric cross-sections and the element abundances to the values provided by \cite{Verner1996} and \cite{Wilms2000}, respectively. When data from different instruments were used simultaneously, a \textsc{constant} component was also included to account for inter-calibration uncertainties between different instruments. We consistently checked that discrepancies between the values for the constant term found for different instruments were not larger than 20\% . When fitting together \nicer and \nustar , a small offset between the slopes of those spectra could be appreciated, as often observed using data from different instruments \citep[see, e.g.,][]{Ludlam2020}. To take into account such an instrumental issue, we left the $\Gamma$ parameter in the \textsc{Xspec} Comptonisation component untied between \nicer and \nustar , in order to allow some flexibility in between the two data sets. However, we checked systematically that the difference in $\Gamma$ was never above 10\%.

Among the seventeen \nicer\ observations used, several have been performed in a span of a few days within each other. In order to maximize the statistics in each spectrum, we checked whether any difference in flux and/or hardness could be appreciated between close-in-time observations. If the spectra were compatible, we summed them by means of the FTOOL \textsc{addspec}. Otherwise, we analysed them separately. This strategy leaves us with eleven final \nicer\ observations. The three \nustar\ observations caught the source during two high modes and a low mode. In all three cases, at least one \nicer\ observation was performed within $\sim$days from the \nustar\ visit, giving us the opportunity to investigate the broadband X-ray spectral behaviour of the source at the extremes of its super-orbital oscillation. In particular we paired observations Ni02 with Nu01, Ni06 with Nu02, Ni10 with Nu03. In the following, these broadband spectra will be simply referred to as A, B and C respectively. 

\subsection{Broadband spectra: the continuum}
An initial fit of spectra A, B and C with a simple absorbed \texttt{powerlaw} model resulted in poor fits in all cases. We replaced the power-law with a more physically motivated thermal Comptonisation model, using the convolution model \texttt{thcomp} \citep{Zdziarski2020_thcomp} and a blackbody model \texttt{bbodyrad} as seed photons spectrum. The involved parameters are the power-law photon index $\Gamma$, the electron temperature of the Comptonising medium $kT_e$, the fraction of the seed source covered by the scattering cloud $f_{\rm cov}$, the blackbody temperature $kT_{\rm bb}$ and its normalisation $K_{\rm bb}$, the latter being connected to the actual blackbody radius of the source through the formula: $K_{\rm bb}=\left(R_{\rm bb}/D_{\rm 10 \ kpc}\right)^2$, with $D_{\rm 10 \ kpc}$ the distance of the source in units of 10~kpc. Using a model with a high energy roll-over such as \texttt{thcomp} improved the fit, but apparent residuals at low energies suggested the presence of an additional spectral component to be included. We therefore added a \texttt{diskbb} component, characterised by a disc blackbody temperature $kT_{\rm disc}$ and a normalisation $K_{\rm disc}$ which, similarly to the \texttt{bbodyrad} normalisation, can be translated into the size of the black body emitting region by the formula $K_{\rm disc}=\left(R_{\rm disc}/D_{\rm 10 \ kpc}\right)^2 \cos{i}$ with $i$ the inclination of the system. In the disc case, the size of the emitting region can be considered as the inner radius of the disc. Using \texttt{diskbb}, the fit to the continuum becomes acceptable  and the residuals are significantly flattened. The model chosen is therefore:

\begin{equation}
    {\rm Model \ 0}: \texttt{tbabs}\times(\texttt{thcomp}\times\texttt{bbodyrad}+\texttt{diskbb})
\end{equation}

We tried to swap \texttt{diskbb} and \texttt{bbodyrad}, testing a scenario where the disc photons serve as Comptonisation seed photons but we found that the resulting estimates for $R_{\rm disc}$ are relatively small for all spectra, in particular for spectrum C with an unphysical upper limit of 8 km. We also tried to apply a second \texttt{thcomp} component to \texttt{diskbb}, driven by the expectation that the corona should also Compton scatter at least a fraction of the disc photons, but without any statistical improvement. \\ 
We used Model 0 also to describe the \astrosat observation (labelled As01), obtaining a satisfactory fit and overall accordance between the best-fit parameters found and A, B and C spectra.

\begin{figure*}
	\includegraphics[width=0.95\textwidth]{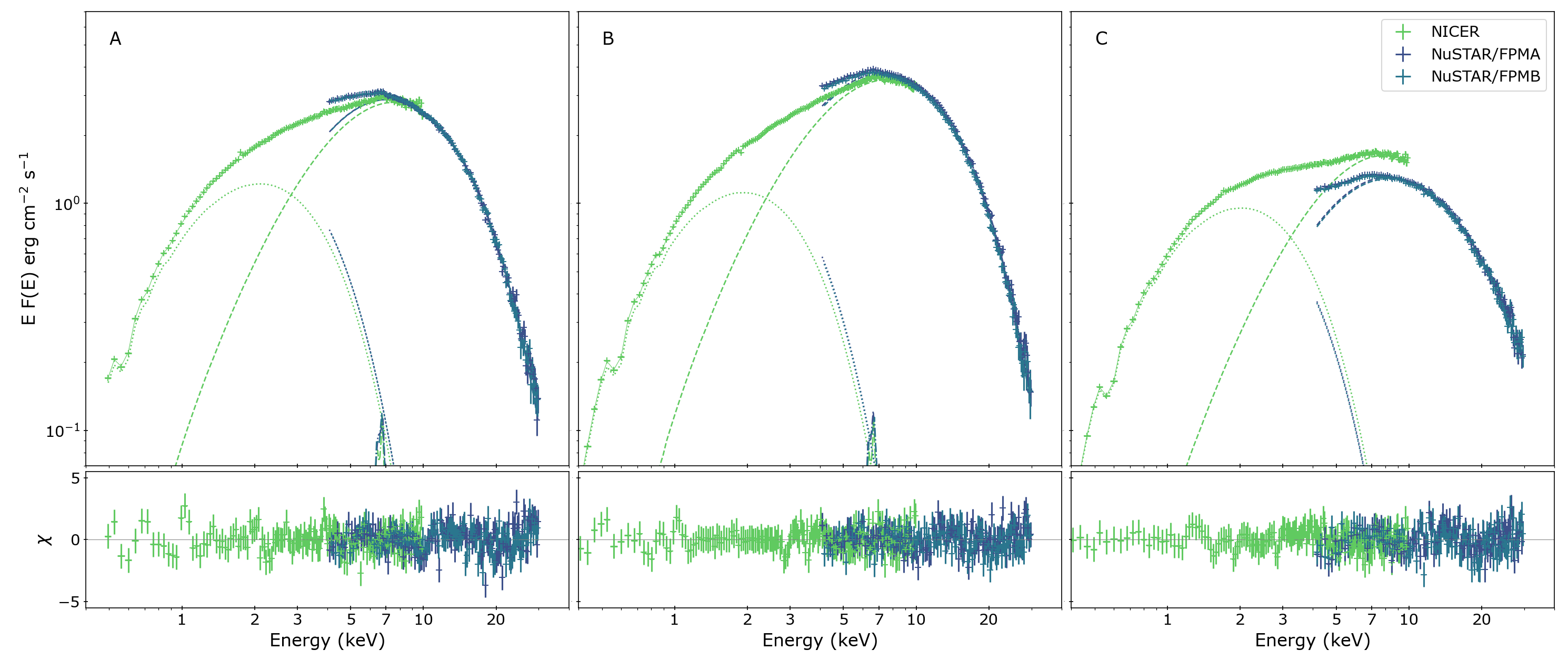}
    \caption{Broadband \nicer (green) and \nustar (blue) spectra with Model 1 and residuals. Different line styles were adopted to distinguish between the different components: dot for \texttt{diskbb}, dash for \texttt{thcomp}$\times$\texttt{bbodyrad} and dash-dot for \textsc{diskline}.}
    \label{fig:ABC_spectra}
\end{figure*}



\subsection{Broadband spectra: discrete spectral features}
Even with a satisfying description of the continuum, some localised residuals could still be spotted at about 6.5-7 keV, clear sign of an unmodelled feature. The presence of residuals in this region has been also reported by other authors, and tackled with a combination of absorption edges \citep[in particular in][]{Mondal2016} or as a reflection Fe line \citep[e.g.,][]{Titarchuk2013}. In this work, we successfully modelled the feature with a \texttt{diskline} component in all three spectra. We refer to the Appendix, section \ref{sec:line}, for a comparison between different models to describe these residuals and to justify our final choice for the \texttt{diskline} model. \\ The line was relatively weak in all observations \citep[as typically observed in UCXBs, e.g.][]{Koliopanos2021}, i.e., normalisation $K_{\rm line}$ ranging from 2$\times$10$^{-3}$ for spectra A and B to 8$\times10^{-4}$ for spectrum C. The contribution to the total flux from the line was about 0.1\%-0.2\% in all cases, so that the lower normalisation in spectrum C can be ascribed to a global fading of the X-ray output from the source. Due to the weakness of the feature, not all line parameters could be well constrained.  We then left the line energy $E_{\rm line}$, the inner radius of the disc $R_{\rm in}$ and $K_{\rm line}$ as the only free parameters, fixing the inclination to 45$^\circ$ \citep[in accordance with the inclination range individuated by][]{Anderson1997}, the emissivity index $\epsilon$ to -2 and the outer radius of the disc $R_{\rm out}$ to 1000 R$_{\rm G}$. The feature is found at an energy $E_{\rm line}$ of $\sim$6.6 keV in all three spectra. Only upper limits could be posed to $R_{\rm in}$, in all cases suggesting a disc that extends relatively close to the NS. 



Residuals in the hard X-ray band were apparently flattened by the simple addition of a \texttt{diskline} and did not show any trace of other reflection features, such as the Compton hump beyond 10 keV. However, the presence of the Iron line feature signals the existence of an underlying reflection component in the spectrum, despite being probably very weak and contributing only marginally to the continuum. A more detailed analysis of the reflection component with self-consistent and more sophisticated models will be presented in a forthcoming companion paper (Anitra et al., in preparation). \\
Clear absorption residuals were also present in the soft X-ray band, below 1 keV. Those features are consistent with O VIII (at $\sim$0.6 keV) and the Ne IX complex (at $\sim$0.9 keV), respectively, which are known to be present in the spectra of the source and have been ascribed to the interstellar medium (ISM) \citep[e.g.][]{Costantini2012}. We improved the fit by multiplying the whole model by two \texttt{gabs} components, i.e., to account for absorption lines of gaussian profiles. \\
With these additions, our final model, herafter Model 1, is the following:

\begin{equation}
    \texttt{tbabs}\times\texttt{gabs}\times\texttt{gabs}\times(\texttt{thcomp}\times\texttt{bbodyrad}+\texttt{diskbb}+\texttt{diskline})
\end{equation}

Furthermore, some absorption features were found in the \nicer spectra at energies of $\sim$1.8 keV, $\sim$2.2 keV as well as an emission feature at $\sim$1.7 keV. The nature of such lines is most likely instrumental, e.g. due to silicon and gold in the detector or the filters. We therefore introduced some additional gaussian features to take them into account. \\
For the fit to the \astrosat As01 spectrum, we included the \texttt{diskline} component, but left out instead the two absorption gaussian component. Indeed, \astrosat known calibration issues makes evaluate the presence of such features more challenging. The broadband spectra fitted with Model 1 and respective residuals are shown in Figure \ref{fig:ABC_spectra}, while we refer to Table \ref{tab:fit_broadband} for the best-fit parameters obtained. 


\begin{table*}
\centering
\begin{tabular}{ l l l l l l}
\hline 
\hline
  \multicolumn{6}{c}{\large Broadband spectral analysis} \T \B \\
\hline
 \multicolumn{6}{c}{{\bf model:} \texttt{tbabs}$\times$\texttt{gabs}$\times$\texttt{gabs}$\times$(\texttt{thComp}$\times$\texttt{bbodyrad}+\texttt{diskbb}+\texttt{diskline})} \\
\hline
 & & A1 & A2 & B & C \T \B \\
 {\bf Parameters} & & Ni02+Nu01 & As01 & (Ni06+Ni07)+Nu02 & (Ni13+Ni14)+Nu03 \\
 \cmidrule{3-6}
$N_{\rm H}$ & ($\times$10$^{22}$ cm$^{-2}$) & $0.161^{+0.003}_{-0.004}$ & $0.237\pm$0.007 & 0.165$\pm$0.002 & $0.165^{+0.002}_{-0.001}$ \T \B \\
$\Gamma$ & & $1.84\pm0.03$ & 2.60$\pm$0.03 & $1.70^{+0.03}_{-0.02}$ & $2.34^{+0.07}_{-0.06}$  \T \B\\
$kT_{\rm e}$ & (keV) & $3.17\pm{0.04}$ & (4.80) & $2.97^{+0.03}_{-0.02}$ & $4.85^{+0.15}_{-0.24}$ \T \B\\
$kT_{\rm bb}$ & (keV) & $1.16^{+0.05}_{-0.03}$ & $1.15\pm$0.04 & $1.04^{+0.05}_{-0.04}$ & $1.46^{+0.04}_{-0.06}$\T \B\\
$R_{\rm bb}$ & (km) & 12.1$\pm$5.0 & 12.2$\pm$5.0 & 16.0$\pm{6.0}$ & 6.0$\pm$2.0 \T \B\\
$kT_{\rm disc}$ & (keV) & $0.80^{+0.03}_{-0.03}$ & $0.57\pm$0.02 & $0.73^{+0.03}_{-0.02}$ & $0.811^{+0.010}_{-0.012}$  \T \B\\
$R_{\rm disc}$ & (km) & 18$^{+7}_{-6}$ & 35$^{+14}_{-16}$ & 22$^{+6}_{-7}$ & 16$\pm$3.0 \T \B\\
$E_{\rm diskline}$ & (keV) & $6.62\pm0.06$ & (6.6) & $6.53^{+0.16}_{-0.08}$ &  6.6$\pm$0.1 \T \B\\
$R_{\rm in}$ & (km) & (20) & (20) & $<$35 & $<$18 \T \B\\
$K_{\rm diskline}$ & ($\times10^{-3}$) & 1.6$\pm$0.4 & (1.8) & 1.9$^{+0.4}_{-0.6}$ & 0.8$\pm$0.2 \T \B\\
$F_{\rm bol}$ & \tiny{($\times$10$^{-8}$ erg cm$^{-2}$ s$^{-1}$)} & 1.21$\pm$0.10 & 1.20$\pm$0.15 & 1.34$\pm$0.15 & 0.70$\pm$0.08 \T \B \\
\cmidrule{3-6}
$\chi^2_\nu$ & (d.o.f.) & 1.02(377) & 1.13(489) & 0.75(390) & 0.93(374) \T \B \\
\hline
\hline
\end{tabular}
\caption{Results of the spectral analysis of our 4 broadband spectra (3 \nicer+\nustar and 1 \astrosat).  Quoted errors reflect 90\% confidence level. The parameters that were kept frozen during the fits are reported between round parentheses. The reported flux values correspond to the 0.01--100~keV energy range.  The values of $R_{\rm disc}$ have been calculated assuming a distance of 7.6 kpc \citep{Kuulkers2003} and an inclination of 45$^\circ$ \citep{Anderson1997}.}
\label{tab:fit_broadband}
\end{table*}

\subsection{Results on the \nicer monitoring}\label{ss:nicer}
In order to reconstruct a more detailed physical evolution of the system during this cycle, we then analysed individually each of the final 11 \nicer spectra (see above for the criteria followed in pairing some of the original 17 spectra to increase the statistics). We consistently used a modified version of Model 1. Indeed, without the hard X-rays coverage provided by \nustar and/or \astrosat , the degeneracies between $kT_e$, $\Gamma$ and the blackbody parameters made the parameters of the fits with \texttt{thcomp}$\times$\texttt{bbodyrad} completely unconstrained. We therefore decided to replace this component with the simpler \texttt{nthcomp}, the main difference being the lack of knowledge on the normalisation of the seed photons spectrum. For the same reason, we had to fix the seed photons temperature in \texttt{nthcomp} $kT_{\rm seed}$ to 1.0 keV, compatible with the values obtained for the broadband fits. In addition, for the modelling of the discrete features, we had to fix some parameters as the fit was unable to find meaningful constraints for them. In particular, we fixed $R_{\rm in}$ in \texttt{diskline} to 20 R$_{\rm G}$ and the energies of the absorption features $E_{\rm line,1}$ and $E_{\rm line, 2}$ to 0.68 keV and 0.87 keV respectively. The best-fit parameters obtained are shown in Tab. \ref{tab:fit_nicer_all}. In Fig. \ref{fig:nicer-all} we display how the main best-fit parameters, the 0.5-10 keV flux and the hardness ratio evolve over time during this cycle. From the Table and the plot it is apparent how some parameters, e.g., $K_{\rm compt}$, seem to follow the super-orbital modulation traced by the flux, while others, in particular the disc temperatures $kT_{\rm disc}$ and radius $R_{\rm disc}$, appear stable.


\section{Timing analysis}\label{sec:timing}
\begin{figure}
\includegraphics[width=0.9\columnwidth]{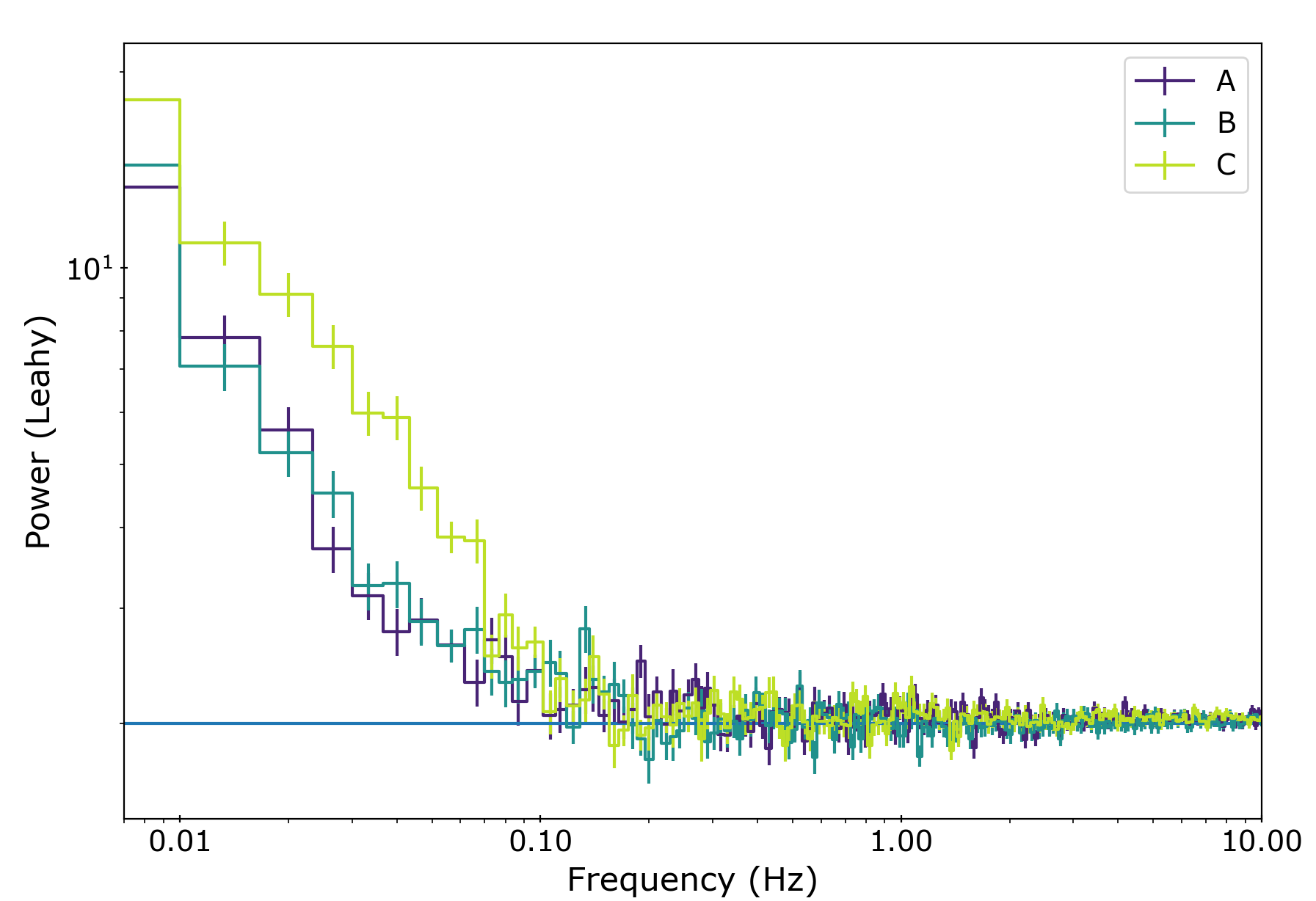}
    \caption{X-rays PDS for the \nustar (3-25 keV) observations used in the broadband spectral analysis. A \textsc{cons} model at value 2.0 is displayed as an horizontal black line.}
    \label{fig:PDS}
\end{figure}

In order to investigate the short-term X-ray variability of the source during the mid-2022 accretion cycle, we extracted Leahy-normalised power density spectra (PDS) from \nustar , \nicer and \astrosat /LAXPC using as energy ranges 3-25 keV, 0.5-10 keV and 3-25 keV, respectively. We performed dead-time correction on each \nustar PDS using the Fourier Amplitude Difference (FAD) technique \citep{Bachetti2018} and then extracted PDS averaging over 150-s long segments with bin time of 1 ms. A representative sample of the obtained PDS for \nustar is shown in Figure \ref{fig:PDS}. For \nicer we broke down the observations in 26-s long segments and used a bin-time of 0.2 ms. We then averaged the PDS created from each segment to produce one averaged PDS per observation with a Nyquist frequency of $\approx$ 2.5~kHz. Finally, for \textsc{LAXPC}, light curves created with a time resolution of 10 ms were used to create Leahy-normalized PDS with 1/8192 s time bins. Also in this case, we average the power spectra obtained from all the segments to obtain one resultant power spectrum. Instead of subtracting the Poisson noise contribution, we rather fitted it with a constant component. \\
We did not find any significant detection of discrete features like quasi-periodic oscillations in those spectra. From each PDS, we estimated the fractional root mean square (RMS) variability, listed in Tables\ref{tab:timing} (see Table \ref{tab:rms} for a complete breakdown of the \nicer sample). While the obtained values for RMS in \nicer show little-to-no evolution through the 17 considered snapshots, the third \nustar observation (C) shows a stronger RMS variability compared to the previous ones (A-B). 
  

\begin{table*}
    \centering
    \begin{tabular}{ c c c c c c c c}
         \hline
         \hline
         \multicolumn{8}{c}{\large Fractional RMS variability (\%)} \\
         \hline
         \multicolumn{2}{c}{\nustar (3-25 keV)} & & \multicolumn{2}{c}{\nicer (0.5-10 keV)} & & \multicolumn{2}{c}{\astrosat / LAXPC (3-25 keV)} \\
         \hline
         Nu01 & 8.5$\pm$0.6 & & Ni02 & 3.4$\pm$1.0 & & As01 & 5.0$\pm$0.5 \\
         Nu02 & 5.0$\pm$0.9 & & Ni06 & 2.8$\pm$1.7 & & \\
         Nu03 & 13.9$\pm$1.0 & & Ni13 & 2.6$\pm$1.3 & & \\
        \hline
        \hline
    \end{tabular}
    \caption{Fractional RMS variability for the \nicer , \nustar and ATCA observations used in the broadband spectral analysis, computed for the frequency range 0.1-100 Hz. All reported errors and upper limits correspond to a confidence levels of 3$\sigma$.}
    \label{tab:timing}
\end{table*}

\section{Radio observations}
\subsection{ATCA data reduction}
\source\ was observed by the Australia Telescope Compact Array (ATCA) 13 times throughout 2022. In the present work, we will only use the observations which were performed close to the \nustar observations, while an analysis of the complete sample will be presented in a forthcoming paper (Russell et al., in preparation). The dates of the three observations presented here are: 2022 April 15, May 28 and July 10. During these observations, ATCA was in its more extended 6D, 1.5B, and 6B configurations\footnote{\url{https://www.narrabri.atnf.csiro.au/operations/array_configurations/configurations.html}}, respectively. In all cases, the fixed location antenna 6 (located 6\,km from the array core) was used during the analysis, providing angular resolutions of $\sim$a few arcseconds for all observations. Observations were recorded simultaneously at central frequencies of 5.5 and 9\,GHz, with a bandwidth of 2\,GHz at each central frequency comprised of 2048 1-MHz channels. We used PKS B1934$-$638 for bandpass and flux density calibration, and B1817$-$254 for phase calibration. Data were flagged, calibrated, and imaged following standard procedures\footnote{\url{https://casaguides.nrao.edu/index.php/ATCA_Tutorials}} in the Common Astronomy Software Application (\textsc{casa} version 5.1.2; \citealt{Casa2022}). Imaging used a Briggs weighting scheme with a robust parameter of 0, balancing sensitivity and resolution, providing angular resolutions of $\sim$a few arcseconds. 

\subsection{Results}
The flux density, $S_{\nu}$ at frequency $\nu$, of the point source was measured by fitting an elliptical Gaussian with full width at half maximum (FWHM) set by the synthesised beam shape. Errors on the absolute flux density scale include conservative systematic uncertainties of 4\% for the 5.5/9\,GHz ATCA data\footnote{\url{https://www.atnf.csiro.au/observers/memos/d96783~1.pdf}} \citep[e.g.,][]{2010MNRAS.402.2403M}, which were added in quadrature with the root mean square (RMS) of the image noise. The radio luminosity, $L_{\rm R}$, was calculated as $L_{\rm R} = 4 \pi S_{\nu} \nu D^{2}$, where $\nu$ is the observing frequency and $D$ is the distance to the source. \\
According to our measurements, a significant radio enhancement can be appreciated during the low X-ray mode (compared to the high-mode). In particular, the radio flux density increases from 70-110 $\mu$Jy (in epochs A and B) to $\sim$500 $\mu$Jy in epoch C. This is the brightest radio flux density recorded for \source at these frequencies \citep[see, for records of radio flux density of the source,][]{Migliari2004,DiazTrigo2017,Russell2021}. \\
We then estimated the radio spectral index $\alpha$, where $S_{\nu}\propto\nu^{\alpha}$, in order to explore the properties of the outflow in each of the three observations. A flat/inverted radio spectrum, where $\alpha\gtrsim 0$, is associated with persistent synchrotron emission from an optically-thick, self-absorbed compact jet, typically observed in XRBs during their hard states. On the other hand, steep radio spectra, where $\alpha \approx -0.7$, are observed from discrete, optically-thin knots of synchrotron emitting plasma that are launched from the system, often referred to as transient jets. Transient jets are detected as the source transitions from the hard to soft state \citep[e.g.][]{Fender2001}. \\
For \source the radio spectral shape was found to be flat during epoch C ($\alpha\sim$-0.2), steep for epoch B ($\alpha\sim$-0.9), and unconstrained for epoch A, being consistent with flat, inverted or steep. This implies a dramatic evolution of the jet properties between high and low modes, as already reported by \cite{Russell2021}. We will compare this intriguing trend with the accretion flow evolution tracked by the X-rays data analysis in Section \ref{ss:jet}.

\begin{table*}
    \centering
    \begin{tabular}{ c c c c c }
         \hline
         \hline
         \multicolumn{5}{c}{\large ATCA observations} \\
         \hline
         Start date & End date & 5.5 GHz Flux density & 9 GHz Flux density & $\alpha$  \T\\
         (UTC) & (UTC) & ($\mu$Jy) & ($\mu$Jy)  \B\\
         \hline
         2022-04-15T18:18:40 & 2022-04-16T00:22:00 & 76$\pm$16 & 74$\pm$16 & -0.1$\pm$0.7 \\
         2022-05-28T13:52:50 & 2022-05-28T19:19:50 & 110$\pm$13 & 74$\pm$11 & -0.9$\pm$0.4 \\
         2022-07-10T12:43:30 & 2022-07-10T18:01:40 & 500$\pm$23 & 450$\pm$25 & -0.22$\pm$0.15 \\
        \hline
        \hline
    \end{tabular}
    \caption{Results from our ATCA radio observations. A bandwidth of 2\,GHz is associated to each frequency band. Flux density errors include systematic uncertainties. Radio spectral indices, provided as $\alpha$, i.e., where $S_{\rm \nu} \propto \nu^{\alpha}$, are also reported.}
    \label{tab:radio}
\end{table*}

\begin{figure}
\includegraphics[width=1.0\columnwidth]{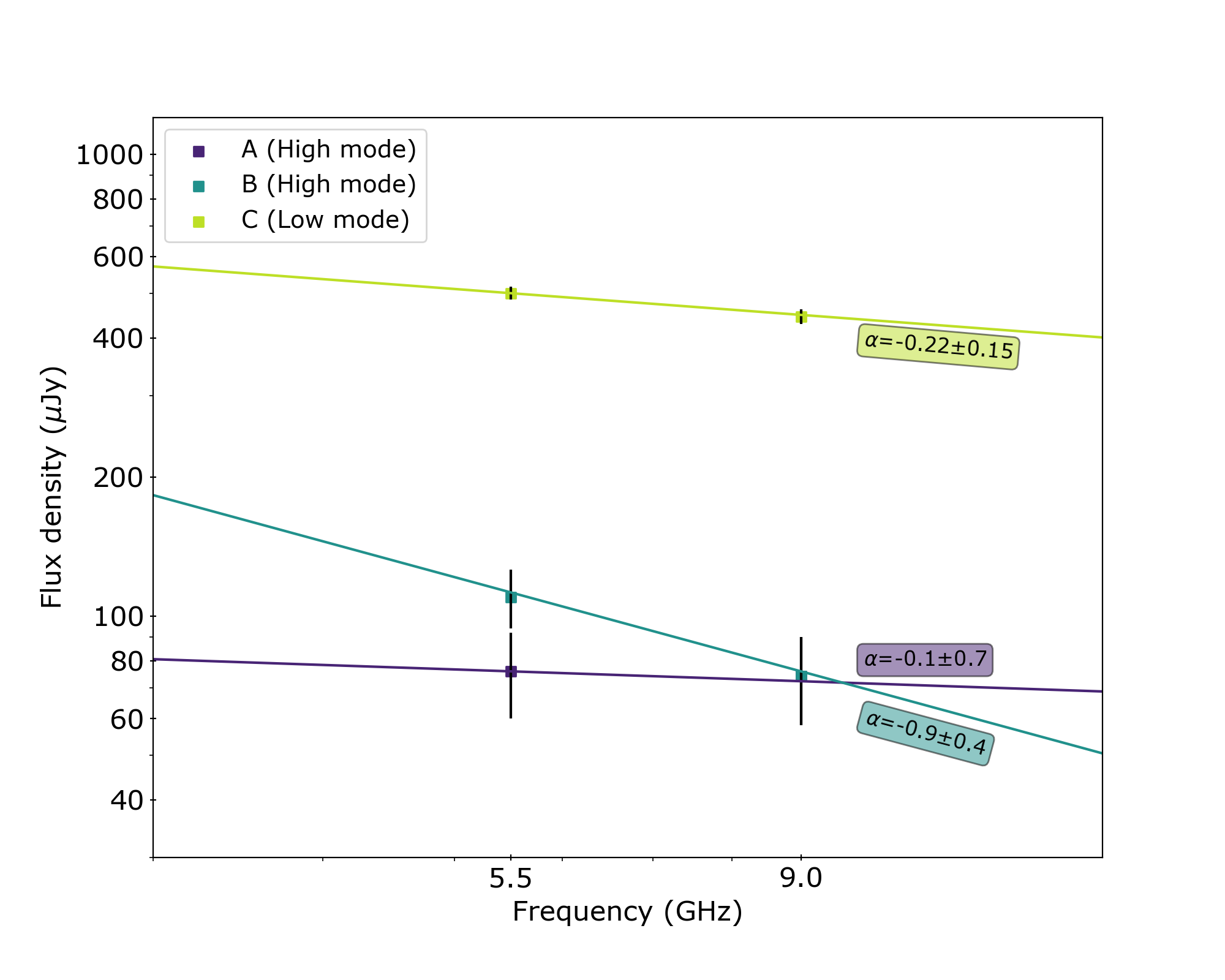}
    \caption{Radio emission from \source during epochs A, B and C (see Tab. \ref{tab:radio} for more details). Best-fit models for the spectra of each epoch are shown. Radio spectral indices $\alpha$ are reported on the plots as well. A dramatic change in radio flux can be appreciated between epoch C (in low mode) and epochs A and B (high mode). In addition, epoch B shows also a steeper spectrum.}
    \label{fig:radio}
\end{figure}

\section{Discussion}\label{sec:disc}
\source provides an almost unique opportunity to observe super-orbital modulations of the accretion rate over relatively short time-scales in a X-ray binary. The spectral evolution of the source during several accretion cycles has been studied in detail by \cite{Titarchuk2013} using Rossi X-ray Timing Explorer, at energies higher than 3 keV. In the work presented here, thanks to the excellent effective area of \nicer\ in the soft X-ray band, we are able to investigate the system's behaviour during one super-orbital modulation in the soft X-ray band for the first time. In addition, we see how the radio emission, and, therefore, the jet, evolved corresponding to these changes in the X-ray spectra. 
\subsection{The origin of the fluctuation in the Comptonisation component and the accretion flow geometry}\label{ss:spreading}
In this paper we showed how the X-ray spectral-timing parameters evolve as the systems bounces between high and low modes. According to the traditional classification scheme based on the colour-colour diagram, \source lingered in the banana state for the whole considered period, moving back and forth between the upper and the lower banana branch. Such a trend seems to reflect more an evolution in the flux rather than in the hardness value,  whose relative stability is consistent with the lack of a spectral transition. We found that a model composed of a disc blackbody, a Comptonisation component and a Fe K line was sufficient to describe the continuum in all the considered spectra. Moving along the super-orbital modulation, however, these two components behave in a rather distinct way. The disc parameters stay more or less constant, in terms of both temperature (about 0.7-0.8 keV), inner radius (20-30 km) and normalisation. On the contrary, the Comptonisation component displays a substantial evolution mostly in its normalisation, i.e., an almost 50\% reduction in the 0.5-10 keV flux going from high to low mode, compared to the less than 10\% drop in the disc component. Such a trend has been already spotted by \cite{Titarchuk2013} using RXTE data, although with a model slightly different than ours. A closer inspection to the obtained best-fit parameters reveals that such an evolution is driven mostly by the normalisation, thereby the physical size, of the source of the seed photons. Such a source has been described with a black body component and associated to the innermost part of the accretion flow. In the high mode, seed photons seem to be radiated by a bigger region with respect to the low modes, i.e., the size going from about 15 km in spectra A-B to about a third of this value in spectrum C. A change can be also observed in the electron temperature of the corona, which seems slightly anti-correlated with the flux, i.e., going from 3 keV (high mode) to about 5 keV (low mode).  While a clear trend is visible in energy spectra, the PDS show less variability. In particular, the fractional RMS is stable at about 2-3.5\% in all \nicer observations. On the other hand, an increase in the power subtended by the PDS, and thereby in the RMS, can be spotted in \nustar PDS going from A-B (RMS$\sim$5-8\%) to C (RMS$\sim$14\%), i.e. with C being significantly more variable. Such a trend is compatible with the increase in temperature of the corona in spectrum C, as hotter coronae are expected to drive stronger X-rays fluctuations. \\

We can piece together all these observed spectral-timing properties into a single interpretation scheme making the following points: (i) in high mode, the mass-accretion rate increases as a consequence of the accretion cycle characterising the system. (ii) The energy fuelled by accretion is dissipated by a $\sim$15 km region in the BL, resulting in more photons and thereby a stronger cooling of the corona. (iii) In low mode, while $\dot{M}$, and therefore the energy supply to the innermost regions of the system, drops, only a smaller, i.e. $\lesssim$10 km, region of the accretion flow remains hot enough to cool the corona by charging it with photons; this region could be a hot spot on the NS surface or a fraction of the BL. Alternatively, the shrinking of the region that provides seed photons for Comptonisation could be explained by invoking a spreading layer covering a larger fraction of the NS surface in high modes than in low modes, once again in response to a changing $\dot{m}$ in the two regimes. (iv) With less photons injected, the corona cools down less and stays hotter, driving more power in the X-rays PDS.   \\
Recently, the new X-ray spectro-polarimetry mission \emph{Imaging X-ray Polarimetry Explorer (IXPE)} \citep{Weisskopf2022} has opened a new avenue to study the accretion flows in XRBs. An \ixpe observational campaign has been performed between 2022 and 2023 on \source \citep{DiMarco2023}. The source was found in high mode in all those observations. The model used by these authors to analyse the broadband X-rays spectra, consisting of a disc blackbody, a Comptonisation spectrum and a gaussian component for the Fe K line, is consistent with ours. In addition, quasi-simultaneous radio observations (with ATCA ) revealed a steep radio spectrum as in our epoch B. These results are consistent with compact jet quenching or transient ejecta during the high modes of \source . In addition, these authors measured a significant ($\sim$10\%) polarisation degree beyond 7 keV along the direction perpendicular to the disc and interpreted these results as possibly due to a mildy relativistic outflow or to reflection. Further studies are necessary to investigate whether such an outflow can be connected with the ejecta responsible for the radio emission in the same state.

\subsection{A boundary layer-jet (anti-)coupling?}\label{ss:jet}
\begin{figure}
	\includegraphics[width=1.0\columnwidth]{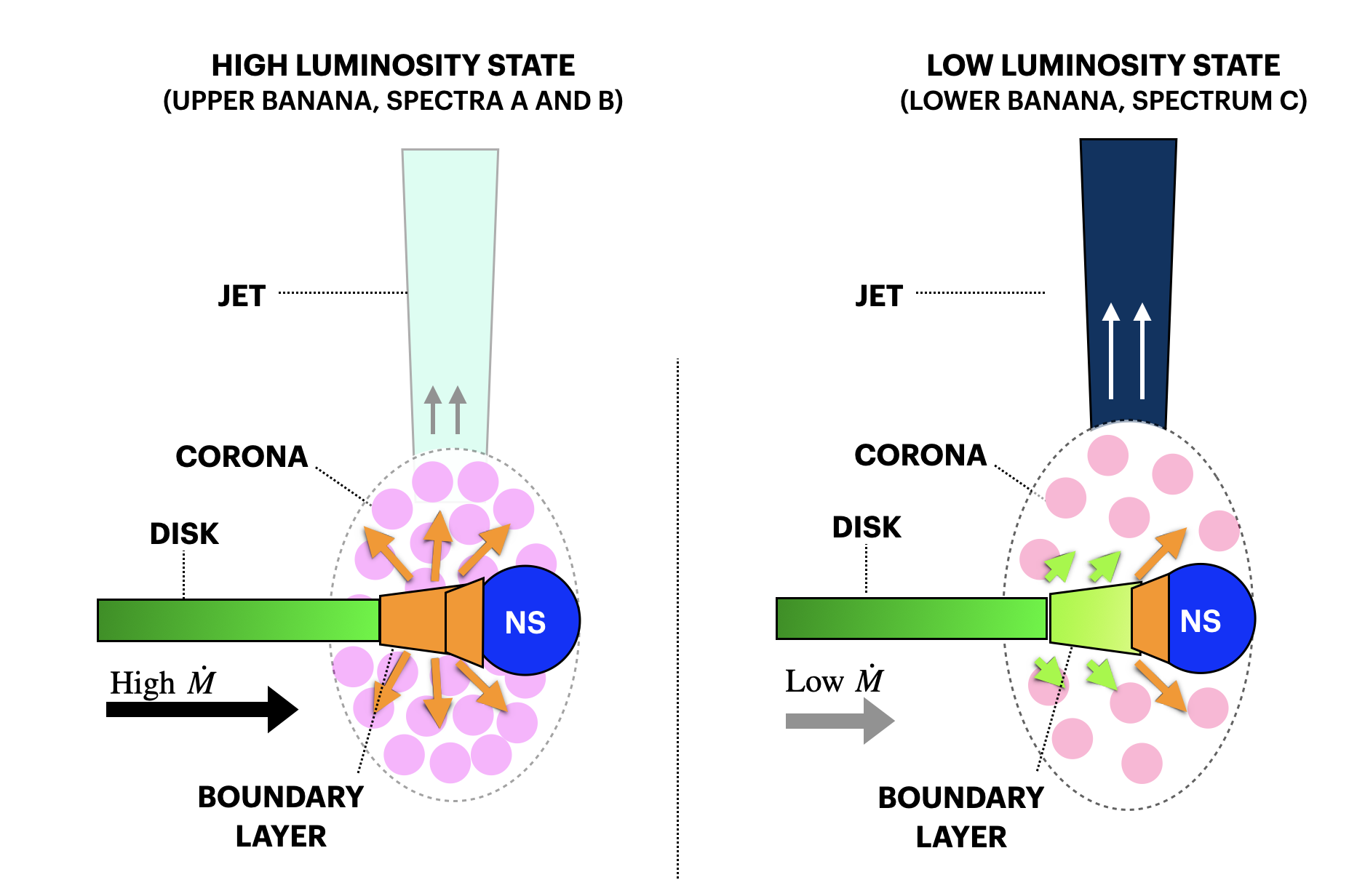}
    \caption{A simplified scheme for our interpretation of the X-rays and radio data. In particular, we show here how the accretion flow and the jet change going from high to low modes.}
    \label{fig:interpretation}
\end{figure}

In the radio band, which is dominated by the jet emission, the source shows dramatic changes between our observations. In the low mode (epoch C), its radio flux density is enhanced by a factor of 5 with respect to the high mode (epochs A and B) and also its spectral shape seems to change. The mode dependency of the jet in \source has already been reported  \cite{Russell2021}, using new and also archival radio data \citep[i.e., from][]{DiazTrigo2017, Panurach2021}, and recently confirmed by \cite{DiMarco2023}. \\ 
In this work, for the first time, we can compare the jet evolution with the simultaneous X-rays spectral-timing properties of the source. In Section \ref{ss:spreading} we have showed that the X-rays evolution during a super-orbital cycle in \source is connected to fluctuations in the boundary layer emission, going from bright to faint in the high to low mode transitions. Such a behaviour produces oscillations in the X-ray light curves that however seem not to be accompanied by transitions between hard and soft states. Indeed, during all of our observations, the source never departs from the vertical "banana" track, as displayed in Fig. \ref{fig:HID}. Despite slight increases in $kT_{\rm e}$ and the hard X-rays RMS seem to indicate that in epoch C a "micro-transition" towards harder states might be ongoing, the spectral and timing properties are not compatible with a full hard state either. Indeed, when \source has been observed in hard / "island" state, it showed remarkably higher $kT_{\rm e}$ \citep[up to 20-30 keV, e.g.,][]{Titarchuk2013} and hard X-rays RMS \citep[up to 20\% and above, e.g.,][]{MunozDarias2014}. Since the observed jet evolution is not occurring in tandem with a soft-to-hard tradition, the behaviour shown by \source seems markedly distinct from what is typically observed in BH and several NS LMXBs \citep[e.g., ][]{Migliari2006,MillerJones2010, Rhodes2022, Fijma2023}. What then determines the evolution in the jet? A link with the boundary layer \citep[which was also suggested by][for the same source]{Russell2021} seems plausible, as it drives the entire evolution of the accretion flow that we observe in our X-ray observations of \source.  According to several accretion/ejection models \citep[e.g.][]{BlandfordPayne1982,Marcel2018a}, jet launching requires the presence of a hot, thin and extended corona \citep[however, see][for observational evidences of an anti-correlation between radio bright jets and hot coronae in the BH XRB GRS 1915+105]{Mendez2022}. The temperature and the geometrical thickness of the corona change as a consequence of the interaction with the photons coming from the disc or, in NS systems, from the BL and the NS surface itself. During the high modes, the boundary layer is bright and extended and irradiates the corona more than during the low modes, making it colder and thicker. On the other hand, during low modes, the corona is charged with less photons and can therefore expand and heat up, an ideal condition to support robust matter ejections. This is also witnessed by the increment in $kT_{\rm e}$ and RMS variability (in \nustar ) going from A/B to C (see Tab. \ref{tab:fit_broadband} and \ref{tab:timing}). Furthermore, it is noteworthy that according to the Internal Shocks Model \citep[]{Malzac2013,Malzac2014}, an increase in X-rays variability would indeed produce brighter jets. Indeed, with more X-rays power we expect larger variability in the velocity with which ejecta in the jet are launched and subsequently more energy dissipated in the shocks between shells moving at different velocity. \\
The proposed scenario is sketched in Fig. \ref{fig:interpretation}. Ultimately in this source, the BL seems to take the main role in regulating the jet properties, to such an extent that we should talk about BL-jet coupling instead of disc-jet coupling, a common term used in the context of LMXBs. An intriguing implication of this scenario is that jet quenching can occur beyond a certain X-ray luminosity but within the same spectral state. Disentangling jet suppression from spectral state transition could be key to explain the proposed presence of compact jets even after the transition to the soft state in a few NS LMXBs \citep[][]{Migliari2004, Migliari2011}. It is also noteworthy that a BL-jet coupling has been already proposed for a system very different from \source , i.e., in the dwarf nova SS Cyg where jet launching was found to be possibly connected with the formation of the boundary layer \citep{Russell2016}. Unfortunately, the radio--X-ray observational campaign presented in this paper is one of the very few available multi-wavelength data sets  of a NS LMXB followed through different regimes and/or spectral states. New radio--X-rays observational campaigns, dense enough to investigate how accretion and ejection evolve over time, have to be performed in the future in order to confirm the existence of such BL-jet coupling in other NS LMXBs. \\
Based on the results presented in this paper, we can speculate that BLs, a rather common ingredient in NS LMXBs, may play a role in regulating jet launching in several different classes, e.g., in in Z-sources and/or in bright, persistent atolls. All of these sources exhibit rather soft spectra, with contribution from both the disc and the BL \citep[see, e.g.][]{Dai2010,Mazzola2021,Saavedra2023}. In addition, Z-sources continuously traverse along three branches in their color-color diagrams, showing drastic changes in both the radio emission and the X-rays variability, but without traditional hard-to-soft transitions \citep{Penninx1988,Migliari2006,Soleri2009}. A similar behaviour has been exhibited also by the bright atoll GX 13+1 \citep{Homan2004}. Finally, jet-related radio emission from the persistent atoll Ser X-1 has been reported while the source was in a soft state \citep{Migliari2004}. \\
On the other hand, we do not expect BLs to form in other classes of NS LMXBs, namely the sources where the disc is truncated very far away from the NS surface. This could be the case for faint atolls typically found in rather hard states, such as Accreting Millisecond X-ray Pulsars \citep[AMXPs][]{DiSalvo2020}. In these sources, the magnetospheric pressure could disrupt the accretion flow in proximity of the compact object, inhibiting the formation of a BL \citep{Degenaar2017, Bult2021,Marino2022}. Of note is that AMXPs can be remarkably radio bright \citep{Russell2018}, sometimes as much as BH LMXBs at the same X-ray luminosity, where BLs can not form either. As an extra ingredient that however may not be present in all NS LMXBs, the role of the BL could then be crucial to explain e.g. why NS LMXBs are typically radio fainter than BH LMXBs, why compact jets could still survive in NS LMXBs during low luminosity soft states and what produces large scatter in radio luminosity within the NS LMXB population \citep[e.g., ][]{Tetarenko2016}. \\


\section{Conclusions}

In this paper, we have presented the results of the 2022 multi-wavelength campaign on the ultra-compact X-ray binary \source , using X-rays (\nicer , \nustar and \astrosat) and radio (ATCA) observatories. Throughout this campaign, we were able to follow how the spectral/timing behavior of the accretion flow and the jet properties evolve in tandem, something rarely seen in NS XRBs. During the surveyed period, the source went through its typical low and high modes, showing a X-ray flux oscillation amplitude of about 40\%. Despite such a strong variability, the source lingered in the banana state and did not show dramatic spectral changes throughout this cycle. A model including a Comptonisation spectrum (with seed photons provided by a boundary layer), the disc multi-colour blackbody emission and a Fe K line worked well in describing the broadband \nicer+\nustar and \astrosat spectra and the single \nicer observations as well. We showed that among these components, the Comptonisation spectrum shows the most dramatic variability, while the disc component is relatively stable. The results of our analysis suggest that modulations in the amplitude of the Comptonisation spectral component are indeed driving such variability, following the very same pattern observed in the X-ray light curve. These modulations suggest an energy flow going from a more extended region in high mode, i.e., the boundary layer, from a more confined region in the low mode, i.e., the NS surface. Furthermore, the hot corona also changes in response to this trend, becoming colder in the high mode as more radiation is pumped into it by the innermost regions of the accretion flow. A correlation between the corona temperature and the values found for the \nustar RMS variability can also be noted. Meanwhile, the jet evolved dramatically in the three observations presented here, brightening considerably during the low mode, coupled with a change in the radio spectrum. We propose that the jet is responding to the changes in the corona, becoming respectively stronger (weaker) as the corona becomes hotter (colder). Ultimately, we suggest that the jet evolution is then driven by the radiation emitted from the BL in an anti-correlation pattern where the jet switches on as the BL switches off. Such a BL-jet (anti-)coupling could be acting also in other NS LMXBs and might in part explain the complex phenomenology of matter ejection in accreting NSs. In order to confirm the above scenario and pinpoint the exact moment where the jet properties evolve, a more detailed look at the radio evolution during this cycle is required (Russell et al., in preparation).




\section*{Acknowledgements}

We thank the anonymous referee for their helpful comments. AMarino, FCZ and NR are supported by the H2020 ERC Consolidator Grant “MAGNESIA” under grant agreement No. 817661 (PI: Rea) and National Spanish grant PGC2018-095512-BI00. This work was also partially supported by the program Unidad de Excelencia Maria de Maeztu CEX2020-001058-M, and by the PHAROS COST Action (No. CA16214). MDS and TDR acknowledge support from the INAF grant "ACE-BANANA". AB is grateful to the Royal Society, United Kingdom. She is supported by an INSPIRE Faculty grant (DST/INSPIRE/04/2018/001265) by the Department of Science and Technology, Govt. of India. A.B. also acknowledges the financial support of ISRO under the AstroSat Archival Data Utilisation Programme (No.DS-2B-13013(2)/4/2019-Sec. 2). TDS and AS acknowledge financial support from PRIN-INAF 2019 with the project "Probing the geometry of accretion: from theory to observations" (PI: Belloni). FCZ is supported by a Ram\'on y Cajal fellowship (grant agreement RYC2021-030888-I). EA acknowledges funding from the Italian Space Agency, contract ASI/INAF n. I/004/11/4. FC acknowledges support from the Royal Society through the Newton International Fellowship programme (NIF/R1/211296). JvdE acknowledges a Warwick Astrophysics prize post-doctoral fellowship made possible thanks to a generous philanthropic donation, and was supported by a Lee Hysan Junior Research Fellowship awarded by St. Hilda’s College, Oxford, during part of this work. We thank Jamie Stevens and ATCA staff for making the radio observations possible. ATCA is part of the Australia Telescope National Facility (https://ror.org/05qajvd42) which is funded by the Australian Government for operation as a National Facility managed by CSIRO. We acknowledge the Wiradjuri people as the Traditional Owners of the ATCA observatory site. NICER is a 0.2--12\,keV X-ray telescope operating on the International Space Station, funded by NASA. NuSTAR is a project led by the California Institute of Technology, managed by the Jet Propulsion Laboratory and funded by NASA. This publication uses the data from the AstroSat mission of the Indian Space Research Organisation (ISRO), archived at the Indian Space Science Data Centre (ISSDC).

\section*{Data Availability}
The X-rays data utilised in this article are publicly available at \url{https://heasarc.gsfc.nasa.gov/cgi-bin/W3Browse/w3browse.pl}, while the analysis products and the ATCA data will be shared on reasonable request to the corresponding author.



\bibliographystyle{mnras}
\bibliography{biblio} 





\bsp	
\label{lastpage}
\appendix
\section{On the modelling of the Fe K line}\label{sec:line}
\begin{figure}
	\includegraphics[width=0.95\columnwidth]{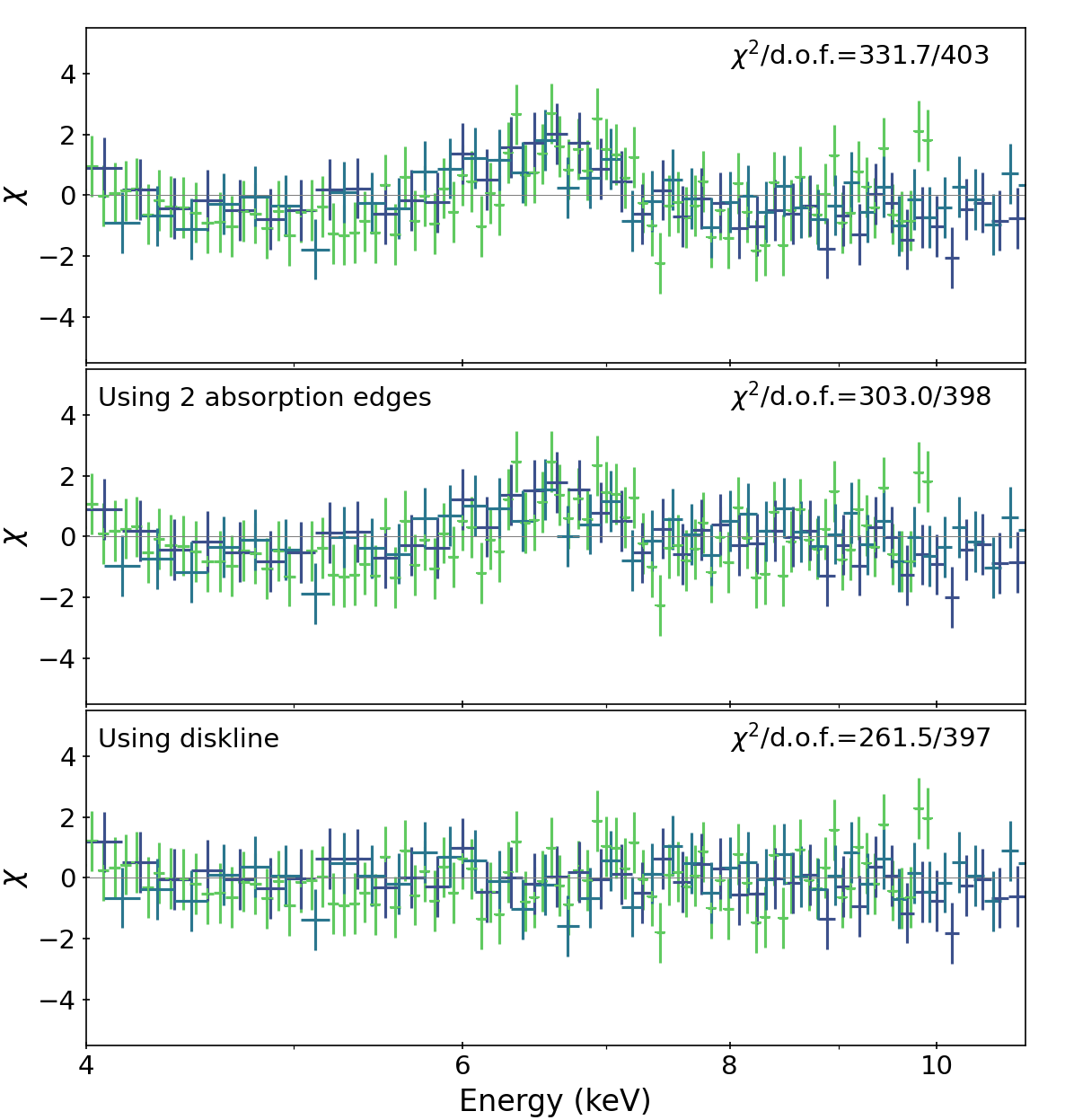}
    \caption{Comparison between the 4--10 keV residuals obtained in Spectrum B with different models to fit the iron line complex region. Models tested: (\emph{Top}) Model 0; (\emph{Middle}) Model 0 plus two absorption edges with energies of$\sim$6.9 keV and $\sim$7.6 keV respectively \citep[following the approach by][]{Mondal2016}; (\emph{Bottom}) Model 0 plus a \texttt{diskline} model. Data: \nicer (green) and \nustar (both FPMA and FPMB, blue).}
    \label{fig:residuals}
\end{figure}

\begin{figure}
	\includegraphics[width=0.7\columnwidth]{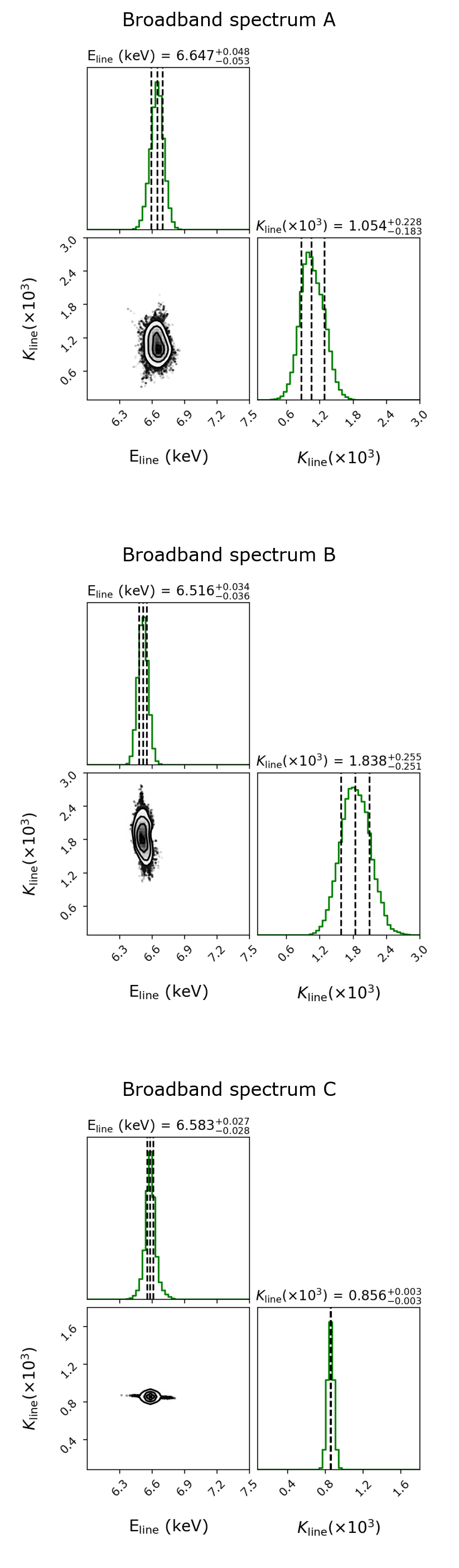}
    \caption{Posterior probability distributions for $E_{\rm line}$ and $K_{\rm line}$ for spectra A (top), B (middle) and C (bottom). Contours represent the 1$\sigma$ , 2$\sigma$ and 3$\sigma$ confidence levels. Marginal posterior distributions are shown as histograms with the median and 1 $\sigma$ intervals of confidence highlighted as dashed lines.}
    \label{fig:contours}
\end{figure}

The 6-7 keV region of all the spectra from \source used in this paper showed residuals, as evident in the top panel of Fig. \ref{fig:residuals}, where we show Spectrum B as a representative example. The presence of similar residuals in this region has been also reported by other authors, and tackled with a combination of absorption edges \citep[in particular in][]{Mondal2016} or as a reflection Fe line \citep[e.g.,][]{Titarchuk2013}. Following the former authors' prescription, we included two absorption edges with the \texttt{edge} model in \texttt{Xspec}. We limited the energies of the edges to the ranges 6.8-7.0 keV and 7.5-7.8 keV respectively, as done for the \nustar spectra in \cite{Mondal2016}. A slight improvement in the fit can be appreciated, but the structure in the residuals is still present (Middle panel, Fig. \ref{fig:residuals}). We therefore removed the edges and added instead a \texttt{diskline} component to the three spectra. \\ Not only does the inclusion of the feature result in a better improvement of the fit, but also in flat residuals. The feature is found significant\footnote{We estimated the significance as the ratio between $K_{\rm line}$ and the error to the normalisation at a confidence level of 1$\sigma$.} in all spectra (significance ranging from 6$\sigma$ to 4.5$\sigma$ confidence). In order to further investigate the line profile, we applied the Goodman-Weare algorithm of Monte Carlo Markov Chain \citep[MCMC;][]{Goodman2010} to produce contour plots for $E_{\rm line}$ and $K_{\rm line}$. We used 20 walkers and a chain length of 5$\times$10$^5$, to calculate the marginal posterior distributions of the best-fit parameters. The results are presented in Figure \ref{fig:contours}, where we use \texttt{corner.py} \citep[][]{Foreman2016} to visualise the MCMC chains.\\ 

\section{A \nicer look at \source}
\begin{figure}
\includegraphics[width=1.0\columnwidth]{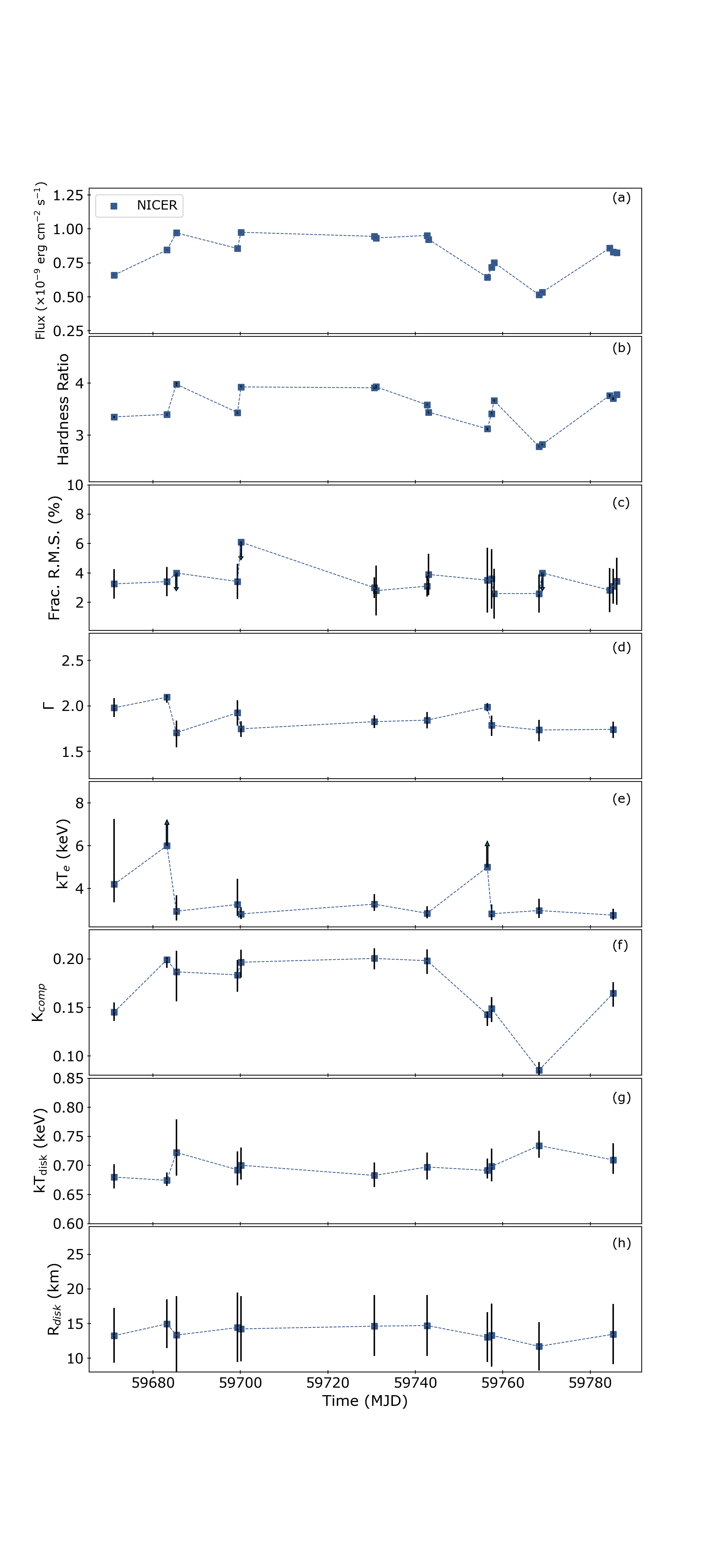}
    \caption{Evolution of the main spectral-timing parameters from the 2022 \nicer campaign. In panel (a) we report the flux in the 0.5-10 keV range as estimated from spectral fitting (see sec. \ref{sec:spectral} for details). Panel (b) shows the \nicer Hardness ratio (2-10 keV/0.5-2 keV). In panel (c) we display the \nicer fractional RMS amplitude in each of the \nicer pointings considered in this work. The evolution of the main best-fit parameters used in the spectral analysis (Sec. \ref{sec:spectral}) is plotted in panels (d)--(i).}
    \label{fig:nicer-all}
\end{figure}

In the following, we present Tables for the spectral and timing analysis performed on any single \nicer observation of this observational campaign (Tables \ref{tab:fit_nicer_all}-\ref{tab:rms}). In order to visualise the changes of the main spectral and timing parameters of the source during this period, we present a tower plot with the time evolution of selected parameters in Fig. \ref{fig:nicer-all}.

\begin{table*}
\centering
\begin{tabular}{ l l l l l l l l}
\hline 
\hline
  \multicolumn{7}{c}{\large \nicer spectral analysis} \T \B \\
\hline
 \multicolumn{7}{c}{{\bf model:} \texttt{tbabs}$\times$(\texttt{nthComp}+\texttt{diskline}+\texttt{diskbb})} \\
\hline
 & 01 & 02 & 03 & 04 & 05 & 06 \T \B \\
{\bf Parameters} & {(Ni01)} & (Ni02) & (Ni03) & (Ni04) & (Ni05) & (Ni06+Ni07)\\
\cmidrule(lr){2-7}
$N_{\rm H}$ \tiny{($\times$10$^{22}$ cm$^{-2}$)}  &  $0.170\pm$0.005 & $0.167\pm$0.004 & $0.167\pm$0.006 & $0.169\pm$0.005 & $0.168\pm$0.005 & $0.168\pm$0.005 \T \B \\
$\Gamma$ &  1.99$\pm$0.11 & 2.10$^{+0.03}_{-0.06}$ & 1.71$^{+0.13}_{-0.16}$ & 1.92$\pm$0.14 & 1.75$^{+0.09}_{-0.08}$ & $1.83\pm 0.07$ \T \B \\
$kT_{\rm e}$ (keV) &  4.2$^{+3.0}_{-0.8}$ & $>$6.0 & 3.0$^{+0.8}_{-0.4}$ & 3.3$^{+1.2}_{-0.5}$ & 2.8$^{+1.3}_{-0.2}$ & 3.3$^{+0.5}_{-0.3}$ \T \B\\
$K_{\rm compt}$ & 0.145$\pm$0.010 & 0.199$^{+0.003}_{-0.009}$ & 0.187$^{+0.030}_{-0.020}$ &  0.184$^{+0.015}_{-0.017}$ & 0.197$^{+0.015}_{-0.013}$ & 0.201$^{+0.011}_{-0.010}$ \T \B\\
$E_{\rm diskline}$ (keV) & 6.53$^{+0.16}_{-0.13}$ & 6.68$\pm$0.10 & $>$6.50 & 6.71$\pm$0.10 & 6.58$^{+0.19}_{-0.12}$ & 6.69$^{+0.15}_{-0.10}$ \T \B \\
$K_{\rm diskline}$ ($\times$10$^{-3}$) & 1.4$\pm$0.5 & 2.1$\pm$0.6 & 1.7$^{+1.4}_{-1.2}$ & 2.2$\pm$0.9 & 1.8$\pm$0.8 & 2.4$\pm$0.7 \T \B \\
$kT_{\rm disc}$ (keV) &  0.68$\pm$0.02 & 0.67$^{+0.13}_{-0.10}$ & 0.72$^{+0.06}_{-0.03}$ & 0.69$\pm$0.03 & 0.69$\pm$0.02 & 0.70$\pm$0.03 \T \B \\
$R_{\rm disc}$ (km) &  13.0$\pm$4.0 & 15.0$\pm$4.0 & 13.0$\pm$6.0 & 14.0$\pm$6.0 & 14.0$\pm$5.0 & 15.0$\pm$4.0 \T \B \\
$\tau_{\rm line, 1}$ (keV) & 0.23$\pm$0.10 & 0.28$\pm$0.10 & 0.24$\pm$0.12 & 0.23$\pm$0.11 & 0.26$\pm$0.11 & 0.26$\pm$0.10 \T \B\\
$\tau_{\rm line, 2}$ (keV) & 0.08$\pm$0.06 & 0.07$\pm$0.06 & $<$0.10 & 0.08$\pm$0.06 &  0.09$\pm$0.06 &  0.09$\pm$0.05 \T \B\\
$F_{X}$ \tiny{($\times 10^{-9}$ erg cm$^{-2}$ s$^{-1}$)} & 6.7$\pm$0.7 & 8.5$\pm$0.9 & 9.7$\pm$1.0 & 8.6$\pm$0.9 & 9.8$\pm$0.1 & 9.4$\pm$0.9  \T \B \\
\cmidrule(lr){2-7}
$\chi^2_\nu$ (d.o.f.) &  0.69(151) & 0.71(142) & 0.83(142) & 0.79(146) & 0.69(154) & 0.66(162) \T \B \\
\\
 & 07 & 08 & 09 & 10 & 11 & \T \B \\
 & (Ni08+Ni09) & (Ni10) & (Ni11+Ni12) & (Ni13+Ni14) & (Ni15+Ni16+Ni17) \\
\cmidrule(lr){2-6}
$N_{\rm H}$ \tiny{($\times$10$^{22}$ cm$^{-2}$)}  &  $0.170\pm$0.005 & $0.168\pm0.005$ & $0.170\pm$0.005 & $0.167\pm$0.005 & $0.165\pm$0.004  \T \B \\
$\Gamma$ & $1.84\pm 0.09$ & 1.99$\pm$0.05 & $1.79^{+0.10}_{-0.12}$ & $1.73^{+0.12}_{-0.11}$ & $1.74\pm0.09$ \T \B \\
$kT_{\rm e}$ (keV) & 2.9$^{+0.3}_{-0.2}$ & $>$5.0 & 2.8$^{+0.5}_{-0.3}$ & 3.0$^{+0.5}_{-0.3}$ & 2.8$\pm$0.3 \T \B \\
$K_{\rm compt}$ & 0.198$^{+0.013}_{-0.012}$ & 0.143$^{+0.012}_{-0.003}$ &  0.149$^{+0.014}_{-0.016}$ & 0.085$^{+0.009}_{-0.008}$ & 0.165$^{+0.014}_{-0.01º}$ \T \B \\
$E_{\rm line}$ (keV) & 6.60$^{+0.15}_{-0.10}$ & 6.59$\pm$0.30 & 6.74$^{+0.15}_{-0.13}$ & $>$6.4 & 6.67$\pm$0.10 & \T \B \\
$K_{\rm line}$ ($\times$10$^{-3}$) & 1.8$\pm$0.6 & 1.0$\pm$0.9 & 1.8$\pm$0.7 & 0.5$\pm$0.4 & 2.0$\pm$0.6 \T \B \\
$kT_{\rm disc}$ (keV) &  0.70$\pm$0.02 & 0.69$^{+0.02}_{-0.01}$ & 0.69$\pm$0.03 & 0.73$\pm$0.03 & 0.71$^{+0.03}_{-0.02}$ \T \B \\
$R_{\rm disc}$ (km) & 15.0$\pm$4.0 & 13.0$\pm$4.0 & 13.0$\pm$5.0 & 12.0$\pm$3.0 & 13.0$\pm$4.0 \T \B \\
 $F_{X}$ \tiny{($\times 10^{-9}$ erg cm$^{-2}$ s$^{-1}$)} & 9.5$\pm$1.0 & 6.6$\pm$0.7 & 7.4$\pm$0.7 & 5.2$\pm$0.5 & 8.4$\pm$0.8 \T \B \\
 $\chi^2_\nu$ (d.o.f.) &  0.61(163) & 0.91(139) & 0.71(149) & 0.65(151) & 0.53(156) \T \B \\
%
\cmidrule(lr){2-6}

\\
\hline
\hline
\end{tabular}
\caption{Results of the spectral analysis of the single \nicer spectra. Quoted errors reflect 90\% confidence level.  The values of $R_{\rm disc}$ have been calculated assuming a distance of 7.6 kpc \citep{Kuulkers2003} and an inclination of 45$^\circ$ \citep{Anderson1997}.}
\label{tab:fit_nicer_all}
\end{table*}

\begin{table*}
    \centering
    \begin{tabular}{ l l l l l l l l l l }
         \hline
         \hline
         \multicolumn{10}{c}{\nicer timing analysis} \\
         \hline
                  & Ni01         & Ni02        & Ni03        & Ni04         & Ni05        & Ni06        & Ni07        & Ni08        & Ni09       \\
         RMS (\%) &  3.3$\pm$1.0 & 3.4$\pm$1.0 & $<$4.0 &  3.4$\pm$1.2 & $<$6.1 & 3.0$\pm$0.7 & 2.8$\pm$1.7 & 3.1$\pm$0.7 & 3.9$\pm$1.4 \\
         \cmidrule{2-9}
                  & Ni10        & Ni11        & Ni12        & Ni13        & Ni14        & Ni15        & Ni16        & Ni17       \\
         RMS (\%) & 3.5$\pm$2.2 & 3.6$\pm$2.0 & 2.6$\pm$1.7 & 2.6$\pm$1.3 & $<$4.0 & 2.8$\pm$1.5 & 3.1$\pm$1.2 & 3.4$\pm$1.6 \\
        \hline
        \hline
    \end{tabular}
    \caption{Fractional RMS for each of the \nicer observations used in this work. The values have been calculated for the 0.5-10 keV energy range and for 0.1-100 Hz as frequency range. All reported errors and upper limits correspond to a confidence levels of 3$\sigma$.}
    \label{tab:rms}
\end{table*}

\end{document}